\documentclass[useAMS,usenatbib,twocolumn]{mn2e}
\usepackage{amsmath}
\usepackage{amssymb}
\usepackage{graphicx}
\usepackage{color}

\newcommand{\apjl}{ApJ} 
\newcommand{\apj}{ApJ}
\newcommand{\apjs}{ApJS} 
\newcommand{\aap}{A\&A} 
\newcommand{\araa}{ARA\&A} 
\newcommand{\mnras}{MNRAS} 

\newcommand{\sovast}{SvA}
\newcommand{\aapr}{A\&A Rev.}

\title[]{Principal Component Analysis of Molecular Clouds: Can CO reveal the dynamics?}
\author[Bertram et al.]{Erik~Bertram$^{1}$, Rahul~Shetty$^{1}$, Simon~C.~O.~Glover$^{1}$, Ralf~S.~Klessen$^{1}$, \and Julia~Roman-Duval$^{2}$ \& Christoph~Federrath$^{3}$\\
$^1$Zentrum f\"ur Astronomie der Universit\"at Heidelberg, Institut f\"ur Theoretische  Astrophysik, Albert-Ueberle-Str.~2, 69120 Heidelberg, Germany\\
$^2$Space Telescope Science Institute, 3700 San Martin Drive, Baltimore, MD 21218, USA\\
$^3$Monash Centre for Astrophysics, School of Mathematical Sciences, Monash University, Vic 3800, Australia
}

\begin{document}

\maketitle

\abstract

We use Principal Component Analysis (PCA) to study the gas dynamics in numerical simulations of typical MCs. Our simulations account for the non-isothermal nature of the gas and include a simplified treatment of the time-dependent gas chemistry. We model the CO line emission in a post-processing step using a 3D radiative transfer code. We consider mean number densities $n_0 = 30, 100, 300\,$cm$^{-3}$ that span the range of values typical for MCs in the solar neighbourhood and investigate the slope $\alpha_\text{PCA}$ of the pseudo structure function computed by PCA for several components: the total density, H$_2$ density, $^{12}$CO density, $^{12}$CO $J = 1 \rightarrow 0$ intensity and $^{13}$CO $J = 1 \rightarrow 0$ intensity. We estimate power-law indices $\alpha_\text{PCA}$ for different chemical species that range from 0.5 to 0.9, in good agreement with observations, and demonstrate that optical depth effects can influence the PCA. We show that when the PCA succeeds, the combination of chemical inhomogeneity and radiative transfer effects can influence the observed PCA slopes by as much as $\approx \pm 0.1$. The method can fail if the CO distribution is very intermittent, e.g. in low-density clouds where CO is confined to small fragments.

\endabstract


\section{Introduction}
\label{sec:introduction}

Assessing the dynamics of molecular clouds (MCs) is crucial for understanding star formation. Besides magnetic fields, gravity, chemistry and the thermodynamics of MCs, turbulence plays a fundamental role in the star formation process \citep[see e.g.,][]{MacLowAndKlessen2004,ElmegreenAndScalo2004,McKeeAndOstriker2007,HennebelleAndFalgarone2012}. Turbulent shocks can create (or disperse) local overdense regions or filaments within MCs that may contract to pre-stellar cores and finally form stars.

The characterization of turbulence in the ISM is necessary to understand the star formation process, the structure of the ISM itself and the distribution of turbulent kinetic energy over different spatial scales. In an extremely influential paper, \citet{Larson1981} noted that the then-available observational data suggested that the velocity dispersion of the gas within molecular clouds was correlated with the size of the cloud. He found that the CO linewidth $\delta v$ scaled with the size $\ell$ of a molecular cloud, as $\delta v \propto \ell^{0.38}$. Later studies \citep[e.g.][]{SolomonEtAl1987} found a slightly steeper relationship, $\delta v \propto \ell^{0.5}$. Furthermore, several studies revealed that MCs in other galaxies \citep[e.g.][]{BolattoEtAl2008} and in the ISM in the Galactic center \citep[e.g.][]{OkaEtAl2001,ShettyEtAl2012} also follow a similar relationship. During the last decades, this linewdith-size relationship has been the focus of several studies \citep[e.g.][]{OssenkopfAndMacLow2002,HeyerAndBrunt2004,HeyerEtAl2009,RomanDuvalEtAl2011}.

There are several established methods for measuring velocity fluctuations as a function of spatial scale, e.g. by measuring the spectral correlation function \citep{RosolowskyEtAl1999}, the velocity channel analysis \citep{LazarianEtAl2000,LazarianEtAl2004} or the $\Delta$-variance method \citep{OssenkopfEtAl2008a,OssenkopfEtAl2008b}. Another well-established method is Principal Component Analysis (PCA), applied in this context by \citet{HeyerAndSchloerb1997,BruntAndHeyer2002a}. PCA is a multivariate mathematical tool that reduces redundant information of a dataset by approximating a large number of statistical variables with a smaller number of significant linear combinations. In this way, PCA can provide a description of turbulent motion by measuring velocity fluctuations as a function of spatial scale. Linear regression then yields a slope of the pseudo structure function obtained from observational measurements \citep[see e.g.,][]{HeyerAndSchloerb1997,BruntAndHeyer2002b,HeyerAndBrunt2004,HeyerEtAl2006,RomanDuvalEtAl2011} or from numerical simulations or analysis of synthetic images \citep[see e.g.,][]{BruntAndHeyer2002a,BruntEtAl2003,FederrathEtAl2010}.

Theoretical efforts are necessary to understand the influence of different effects on the PCA slopes, such as geometric projection \citep[see e.g.,][]{OstrikerEtAl2001,Ballesteros-ParedesAndMacLow2002,ShettyEtAl2010}, the radiation field, optical depth effects \citep{OssenkopfAndKlessenAndHeitsch2001}, density fluctuations \citep{RomanDuvalEtAl2011} or the non-uniform chemical makeup of the gas. Chemical effects may be important for the determination of velocity fluctuations as a function of spatial scale, since the relation between H$_2$ and CO is not linear. Furthermore, CO accurately traces the total gas only in dense regions, and so CO emission may not accurately reveal the properties of the diffuse gas (\citealt{GloverEtAl2010}; \citeauthor{ShettyEtAl2011a}~2011a).

Recent developments have allowed us to use high-resolution, three-dimensional chemistry models in numerical simulations of turbulent interstellar gas \citep{GloverEtAl2010} to get a better insight into how the choice of chemical species as a tracer of gas affects the slope of the pseudo structure function within MCs. Time-dependent chemistry is important for modelling the formation and destruction of molecules in diffuse environments. Radiative transfer calculations are then required in order to convert the simulation output into synthetic emission maps. We focus in this paper on how chemistry and radiative transfer affect the observed slopes of the pseudo structure function in different environments, spanning a range of possible densities representative of MCs. We also compare the theoretical findings with recent observational efforts by \citet{RomanDuvalEtAl2011}.

In section \ref{sec:methodsandsims} we describe the simulations, the radiative transfer post-processing and the PCA technique, which we use to compute the pseudo structure functions. In section \ref{sec:results} we present the results of our PCA of simulations with different mean number densities. We discuss our findings in section \ref{sec:discussion}, and present our conclusions in section \ref{sec:summary}.

\section{Methods and simulations}
\label{sec:methodsandsims}

\subsection{MHD and chemistry simulations}
\label{subsec:sims}

The simulations that we examine in this paper were performed using a modified version of the ZEUS-MP magnetohydrodynamical code \citep{Norman2000,HayesEtAl2006}. We make use of a detailed atomic and molecular cooling function, described in detail in \citet{GloverEtAl2010} and \citet{GloverAndClark2012}, and a simplified treatment of the molecular chemistry of the gas. Our chemical treatment is based on the work of \citet{NelsonAndLanger1999} and \citet{GloverAndMacLow2007}, and allows us to follow the formation and destruction of H$_{2}$  and CO self-consistently within our simulations. Full details of the chemical model can be found in \citet{GloverAndClark2012}. We model the penetration of radiation into our cloud using the six-ray approximation \citep{NelsonAndLanger1997,GloverAndMacLow2007}. In this approximation, we compute photochemical rates in each zone in our simulation volume by averaging over the six lines of sight that lie parallel to the coordinate axes. We include the effects of H$_{2}$ self-shielding, CO self-shielding, and the shielding of CO by H$_{2}$, using shielding functions taken from \citet{DrainAndBertoldi1996} and \citet{LeeEtAl1996}.

Our initial conditions are similar to those already studied by \citet{GloverEtAl2010} and \citet{GloverMacLow2011}. We consider a periodic volume, with side length 20~pc, filled with uniform density gas. We run simulations with three different initial densities, $n_{0} = 30, 100$ and $300 \: {\rm cm^{-3}}$, where $n_{0}$ is the initial number density of hydrogen nuclei. The initial temperature of the gas is set to 60~K. We initialize the gas with a turbulent velocity field with uniform power between wavenumbers $k = 1$ and $k = 2$ \citep{MacLowKlessenBurkertSmith1998,MacLow1999}. The initial 3D rms velocity is $v_{\rm rms} = 5 \: {\rm km \: s^{-1}}$, and the turbulence is driven so as to maintain $v_{\rm rms}$ at approximately the same value throughout the run. Our forcing excites a mixture of solenoidal and compressive modes, which is an intermediate case. However, \citet{MicicEtAl2012} showed that the amount and timescale for H$_{2}$ formation depends on the forcing of the turbulence. \citet{MicicEtAl2012} find that compressive forcing produces H$_{2}$ twice as fast as purely solenoidal forcing, because compressive forcing compresses the gas more efficiently \citep{FederrathEtAl2008}, leading to higher densities and thus faster H$_{2}$ formation \citep{HollenbachWernerSalpeter1971}. Here we only analyse the mixed forcing case and leave the more extreme cases of purely solenoidal and compressive forcing for future studies. The gas is magnetized, with an initially uniform magnetic field strength $B_{0} = 5.85 \mbox{ }\mu {\rm G}$, initially oriented parallel to the $z$-axis of the simulation. We neglect the effects of self-gravity. 

We assume that the gas has a uniform solar metallicity and adopt the same ratios of helium, carbon and oxygen to hydrogen as in \citet{GloverEtAl2010} and \citet{GloverMacLow2011}. At the start of the simulations, the hydrogen, helium and oxygen are in atomic form, while the carbon is assumed to be in singly ionized form, as C$^{+}$. We also adopt the standard local value for the dust-to-gas ratio of 1:100 \citep{GloverEtAl2010}, and assume that the dust properties do not vary with the gas density. The cosmic ray ionization rate was set to $\zeta = 10^{-17} \: {\rm s^{-1}}$, but we do not expect our results to be sensitive to this choice. For the incident ultraviolet radiation field, we adopt the standard parameterization of \citet{Draine1978}.

\subsection{Radiative transfer}
\label{subsec:RADMC}

To compute the intensity of the CO ($J=1 \rightarrow 0$) line in the model MCs, we employ the radiative transfer code RADMC-3D\footnote{www.ita.uni-heidelberg.de/$\sim$dullemond/software/radmc-3d/}. We calculate the population levels through the Large Velocity Gradient approximation \citep{Sobolev1957}, which uses local velocity gradients to estimate photon escape probabilities in each zone of the simulation. The LVG implementation in RADMC-3D is described in \citeauthor{ShettyEtAl2011a}~(2011a). We also ensure that velocities along the LoS are well resolved, using the Doppler Catching feature in RADMC-3D (as described in \citeauthor{ShettyEtAl2011b}~2011b). We perform the radiative transfer calculation of the $J=1 \rightarrow 0$ line for both of $^{12}$CO and $^{13}$CO. We assume that the ratio of $^{12}$CO to $^{13}$CO, $R_{\rm 12/13}$, remains constant throughout the cloud, and that it has the value $R_{12/13} = 50$. In reality, selective photodissociation \citep[see e.g.][]{Visser2009} and chemical fractionation \citep{WatsonEtAl1976} can alter this ratio, but we do not expect these effects to have a major impact on our results as our intensity maps are dominated by emission from the densest regions, where $R_{\rm 12/13}$ is close to our assumed $^{12}$C to $^{13}$C ratio (Sz\H{u}cs et~al., in prep.).

The radiative transfer calculation yields position-position-velocity (PPV) cubes of brightness temperatures $T_B$. Furthermore, we use the simulated position-position-position (PPP) cubes of density and velocity to construct PPV cubes of column density. We construct such PPV cubes from the total gas density, H$_2$ density and CO density. To generate the PPV cubes from the density fields, we sum up all density values in the PPP cube that fall within a given velocity range of width $\delta v \approx 0.07\,{\rm km \: s^{-1}}$. The velocity of each density cell is directly given by the PPP velocity cube. We will refer to these as ``density'' PPV cubes. For those we use a number of 256 velocity channels for the 256$^3$ resolution models. We perform the PCA on both these density PPV data as well as on the ``intensity'' PPV cubes, generated from the radiative transfer code, and compare the results.

\subsection{Principal component analysis}
\label{subsec:PCA}

Principal component analysis (PCA) is a multivariate mathematical tool that reduces redundant information of a dataset by approximating a large number of statistical variables with a smaller number of significant linear combinations. In astronomy it is used to reduce redundant information from spectroscopic image observations, such as observations of MCs. We work in PPV space, since it represents the spectral line mapping of MCs. In this context, PCA provides velocity fluctuations $\delta v_\text{PCA}$ as a function of spatial scales $\delta \ell$, $\delta v_\text{PCA} (\delta \ell)$, similar to a structure function. The details of the application of PCA to spectral observations of MCs are given in, e.g., \citealt{HeyerAndSchloerb1997}, \citealt{BruntAndHeyer2002a}. We assess the slope and scaling coefficient of the PCA pseudo structure function, $\delta v_\text{PCA} \propto \ell^{\alpha_\text{PCA}}$, in the different simulations and compare the results to observational measurements. We note that deriving the real linewidth-size relation requires certain calibration steps to obtain $\delta v$ in Larson's relation from $\delta v_\text{PCA}$ \citep[see e.g.,][]{BruntEtAl2003,RomanDuvalEtAl2011}.

To carry out a PCA, the 3D PPV cube is transformed to a two-dimensional dataset ${\bf T}(x_i, y_i, v_j) = {\bf T}({\bf r}, v) = T_{ij}$, where ${\bf r} = (x_i, y_i)$ are the spatial positions on the sky and $v_j$ the line-of-sight velocity. The indices $i$ and $j$ denote spatial (of number $N_r$) and spectral (of number $N_v$) positions respectively. In a first step we compute the components of the covariance matrix
\begin{equation}
S_{jk} = \frac{1}{N_r N_v} T_{ij}T_{ik}
\end{equation}
by summation over all $N_r$ spatial positions. Second, we solve the equation
\begin{equation}
{\bf S}{\bf u}^{(k)} = \lambda^{(k)}{\bf u}^{(k)}
\end{equation}
where $\lambda^{(k)}$ are the eigenvalues and ${\bf u}^{(k)}$ the corresponding eigenvectors. The eigenvalues describe the magnitude of variance in the data in the direction of its corresponding eigenvector. The index $k$ is an integer ordering number that labels the different eigenvalues in decreasing order. The projection of the eigenvectors ${\bf u}^{(k)}$ onto the PPV cube ${\bf T}({\bf r}, v)$ yields the principal component ${\bf PC}^{(k)}$, ${\bf PC}^{(k)} = {\bf T}^\text{T}{\bf u}^{(k)}$, which is the eigenimage.
The spatial and velocity scales are then determined by computing the autocorrelation function (ACF) for each eigenvector and eigenimage. The $k$th spatial scale corresponds to the spatial scale at which the ACF of the $k$th PC drops by one e-fold. Correspondingly, the $k$th velocity scale is obtained from the one e-fold distance of the ACF of the $k$th eigen-spectrum. To stay consistent with previous publications, we denote the $k$th spatial scale $\delta \ell$ in the following as $l_k$ and the corresponding $k$th spectral scale as $v_k$.

\begin{figure}
\centerline{
\includegraphics[width=1.0\linewidth]{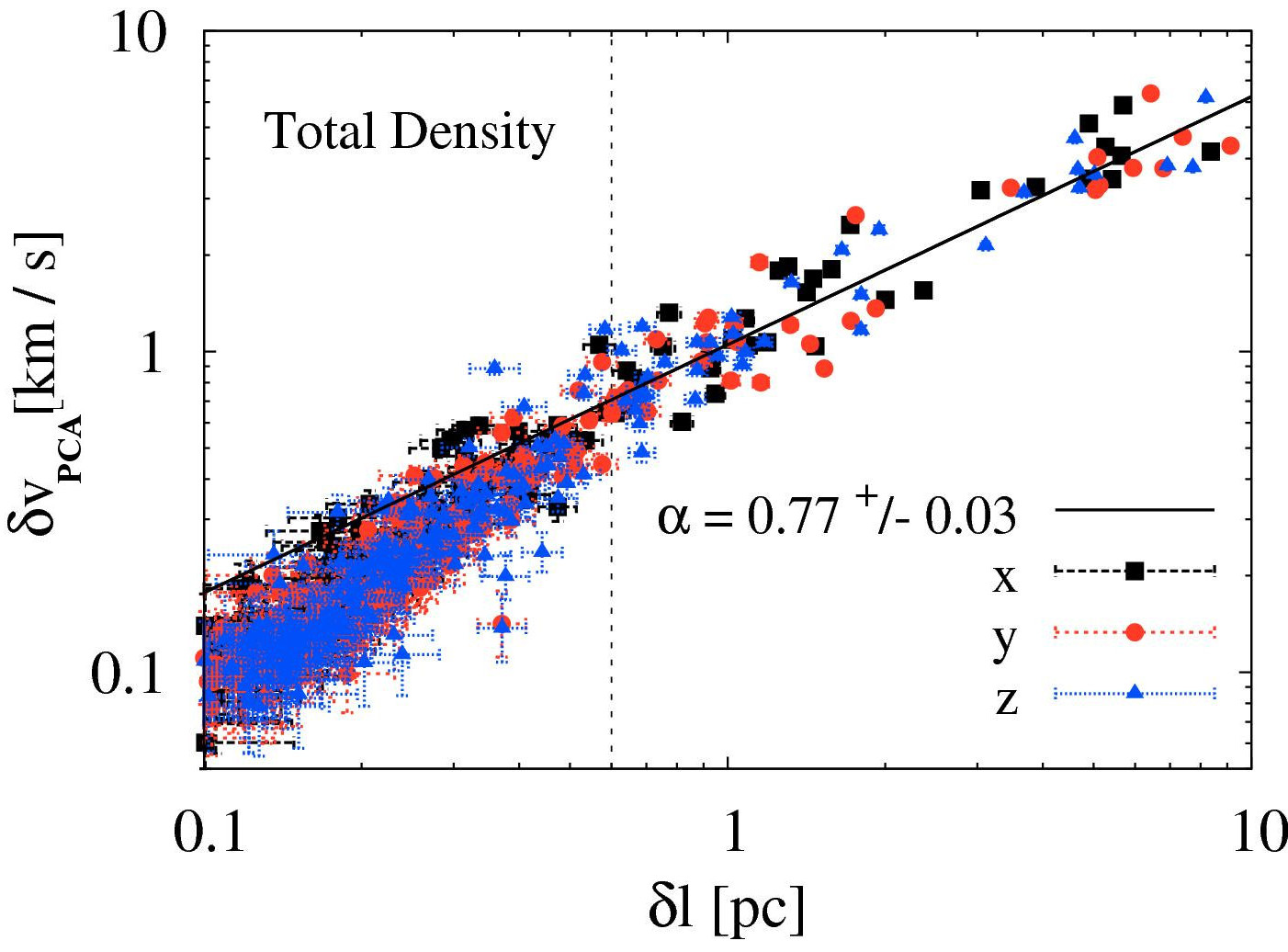}
}
\caption{Velocity fluctuations $\delta v_\text{PCA}$ as a function of spatial scales for the total gas density and different LoS with an initial number density $n_0 = 300\,$cm$^{-3}$. Linear regression gives a PCA slope $\alpha_\text{PCA}$ (listed in the plot) for the pseudo structure function. Shown is the whole spatial range of PCA data available from our analysis. We find an apparent break of the general trend towards scales smaller than \mbox{$\sim0.6\,$}pc (marked with a dashed line). This is due to an inadequate resolution on the smallest scales.} 
\label{fig:turnoff}
\end{figure}

When considering the whole spatial range of the PCA data, there is an apparent break of the general trend towards scales smaller than \mbox{$\sim0.6\,$}pc (see Figure \ref{fig:turnoff}). This is likely due to inadequate resolution on the smallest scales. In addition, as we are only interested in the mid-to-large scale structure, we only evaluate $\alpha_\text{PCA}$ on physical length-scales larger than \mbox{$\sim0.6\,$}pc, corresponding to 8 pixels in the simulation domain for our canonical resolution of $256^{3}$.

For more information about the PCA tool that we use we refer the reader to \citet{HeyerAndSchloerb1997,BruntAndHeyer2002a,RomanDuvalEtAl2011}. Here, we apply PCA to numerical simulations with chemistry and radiative transfer described in section \ref{subsec:sims} and \ref{subsec:RADMC}, and compare our results to existing studies in the literature.

\subsection{Limitations of the model}
\label{subsec:limitations}

There are a number of assumptions and limitations in our simulations that we must bear in mind when interpreting the PCA. For instance, the models do not include self-gravity, star formation, stellar feedback (e.g. by SN, cosmic rays, etc.) and do not account for large-scale dynamics (e.g. spiral arms, SN shells, etc.). There is also a resolution dependence inherent to all numerical simulations of the turbulent cascade (as further discussed in Appendix \ref{app:resol}, see also \citealt{KitsionasEtAl2009,FederrathEtAl2010,FederrathEtAl2011,SurEtAl2010,Federrath2012}) and the chemical network \citep{GloverEtAl2010}. However, in this work we are primarily interested in the comparison between intensity PPV cubes with density PPV cubes, in order to investigate the effects of radiative transfer on the PCA slopes. We are thereby focusing on how reliably we can infer the underlying PCA parameters of the ISM from observational data.

\section{Results}
\label{sec:results}

\begin{table*}
\begin{tabular}{l|c|c|c}
\hline\hline
Model name & n30 & n100 & n300 \\
\hline
Mean density $n_0$ $\lbrack$cm$^{-3}\rbrack$ & 30 & 100 & 300 \\
Resolution & 256$^3$ & 256$^3$ & 256$^3$ \\
$t_{end}$ [Myr] & 5.7 & 5.7 & 5.7 \\
$\langle x_{H_2} \rangle_{mass}$ & 0.35 & 0.69 & 0.94 \\
$\langle n_{H_2} \rangle_{vol}$ $\lbrack$cm$^{-3}\rbrack$ & 7.9 & 35.1 & 138.1 \\
$\langle n_{H_2} \rangle_{mass}$ $\lbrack$cm$^{-3}\rbrack$ & 171.49 & 325.7 & 1015.3 \\
$\langle n_{CO} \rangle_{vol}$ $\lbrack$cm$^{-3}\rbrack$ & $8.33 \times 10^{-5}$ & $1.5 \times 10^{-3}$ & $2.2 \times 10^{-2}$ \\
$\langle n_{CO} \rangle_{mass}$ $\lbrack$cm$^{-3}\rbrack$ & $9.7 \times 10^{-3}$ & $4.3 \times 10^{-2}$ & 0.2 \\
$\langle T \rangle_{vol}$ $\lbrack$K$\rbrack$ & 221.1 & 66.9 & 35.3 \\
$\langle T \rangle_{mass}$ $\lbrack$K$\rbrack$ & 56.1 & 25.7 & 12.8 \\
$\langle N_{H_2} \rangle$ $\lbrack$cm$^{-2}\rbrack$ & $4.9 \times 10^{20}$ & $2.2 \times 10^{21}$ & $8.5 \times 10^{21}$ \\
$\langle N_{CO} \rangle$ $\lbrack$cm$^{-2}\rbrack$ & $5.1 \times 10^{15}$ & $8.9 \times 10^{16}$ & $1.4 \times 10^{18}$ \\
$\langle W_{^{12}CO} \rangle$ $\lbrack$K km s$^{-1}\rbrack$ & 0.9 & 6.6 & 25.5 \\
$\langle W_{^{13}CO} \rangle$ $\lbrack$K km s$^{-1}\rbrack$ & 0.1 & 1.1 & 8.6 \\
\hline
\end{tabular}
\caption{Overview of the different models with some characteristic values. Given are (from top to bottom): mean number density, resolution, time of the last snapshot, mass-weighted mean abundances of H$_2$ \citep[i.e. the percentage of atomic hydrogen beeing converted to H$_2$, see][]{GloverAndMacLow2011}, mean volume- and mass-weighted H$_2$ and CO number density, mean volume- and mass-weighted temperature, mean column density of H$_2$ and CO and mean values of the velocity-integrated intensities of $^{12}$CO and $^{13}$CO.}
\label{tab:setup}
\end{table*}

\begin{figure*}
\centerline{
\includegraphics[height=0.27\linewidth]{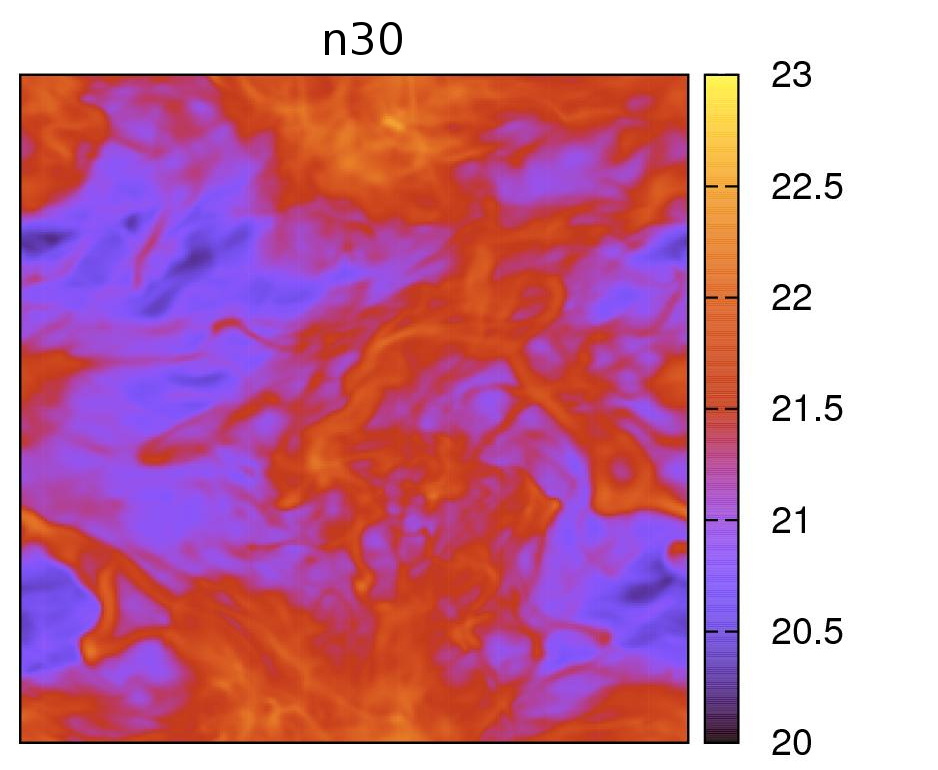}
\includegraphics[height=0.27\linewidth]{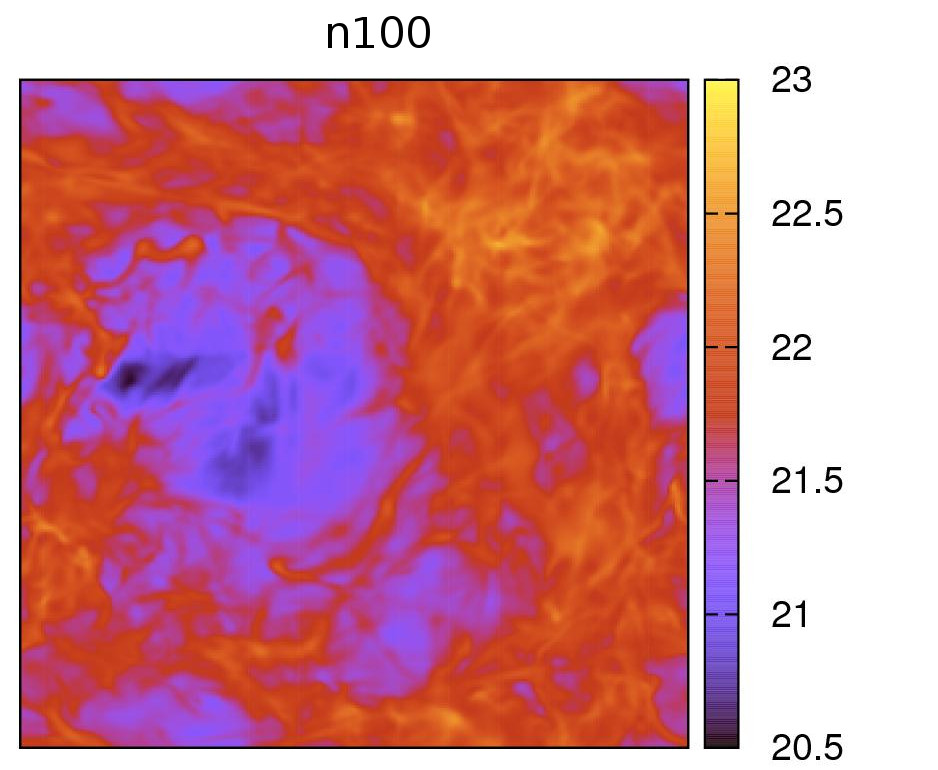}
\includegraphics[height=0.27\linewidth]{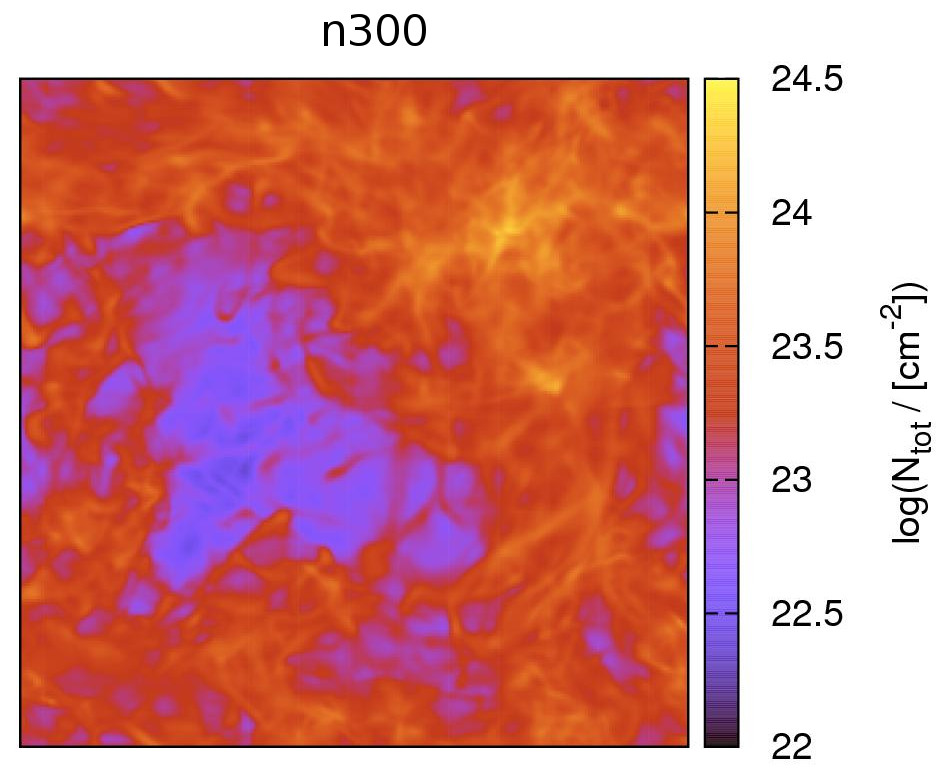}
}
\centerline{
\includegraphics[height=0.25\linewidth]{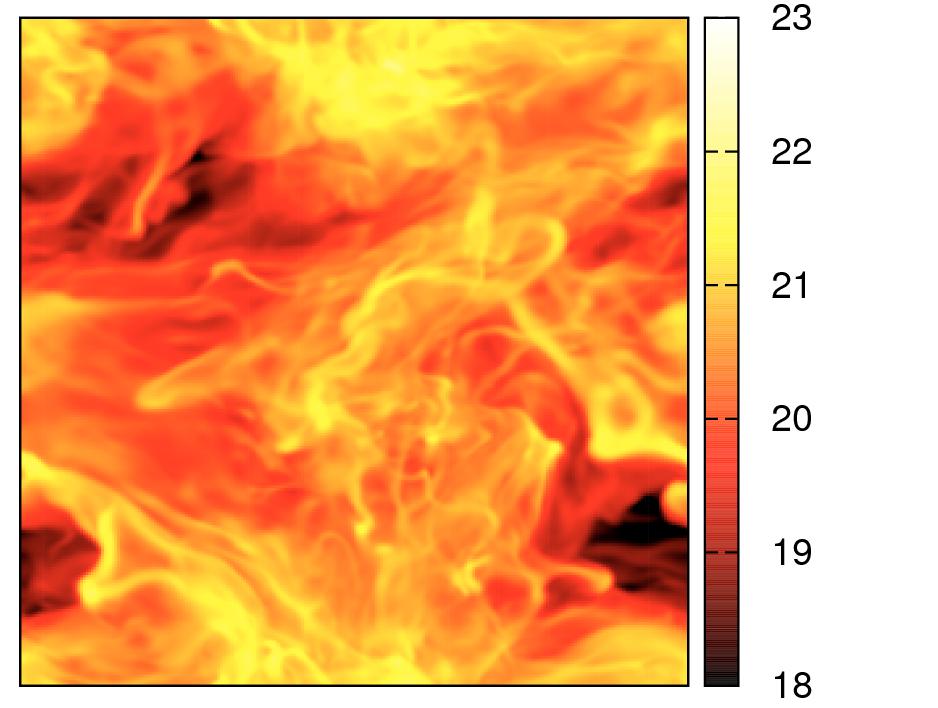}
\includegraphics[height=0.25\linewidth]{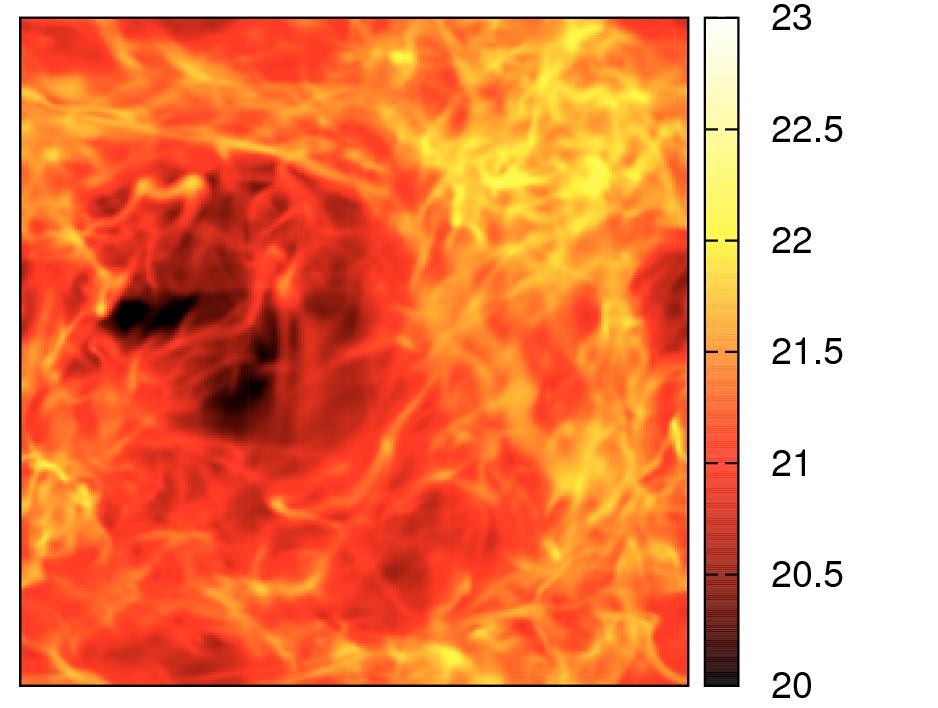}
\includegraphics[height=0.25\linewidth]{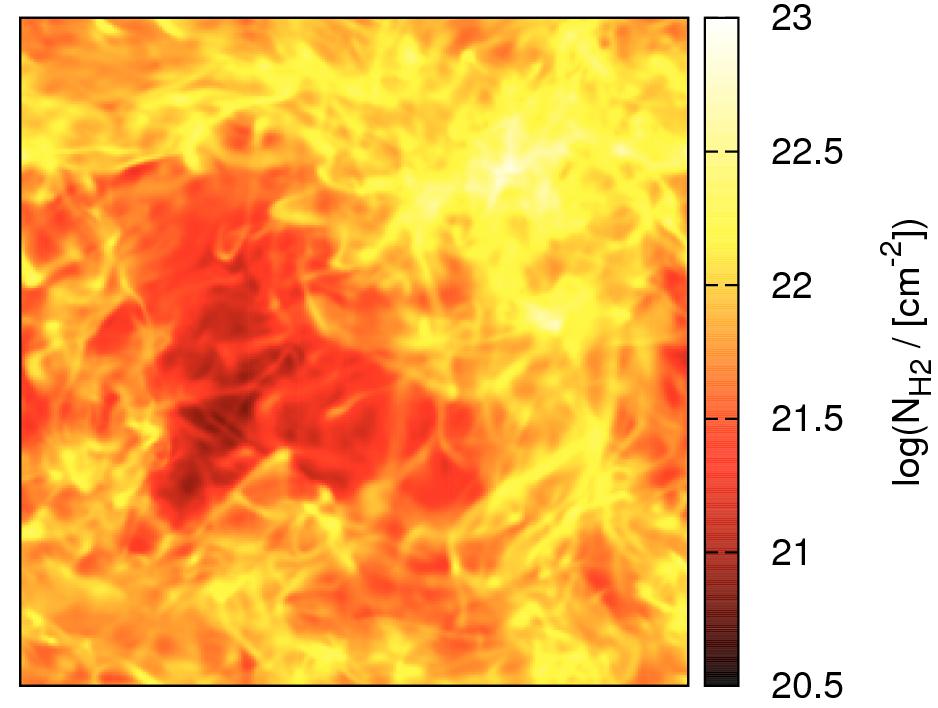}
}
\centerline{
\includegraphics[height=0.25\linewidth]{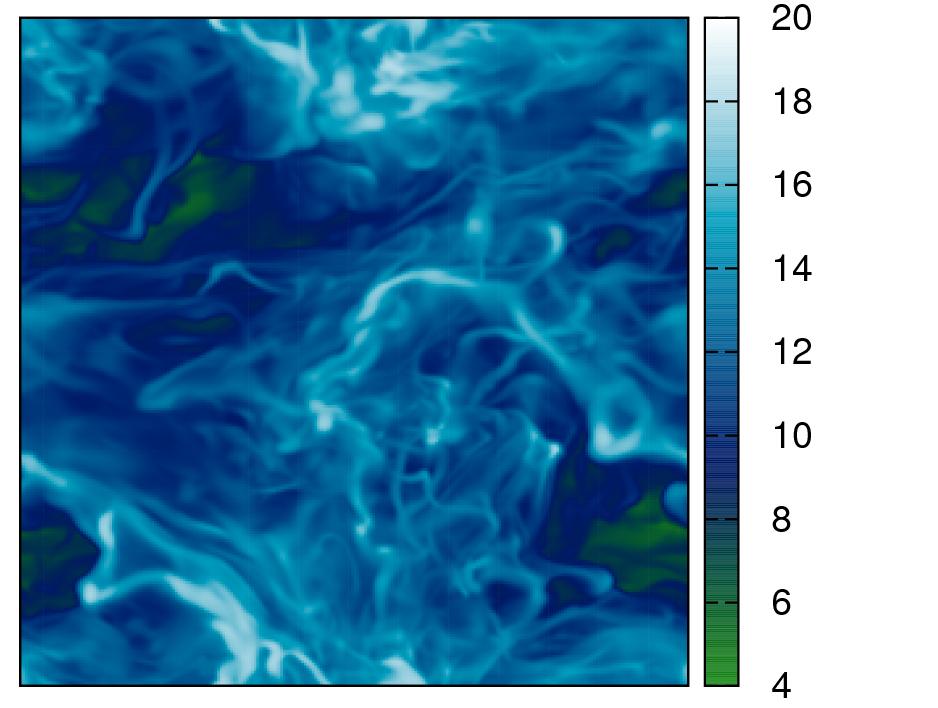}
\includegraphics[height=0.25\linewidth]{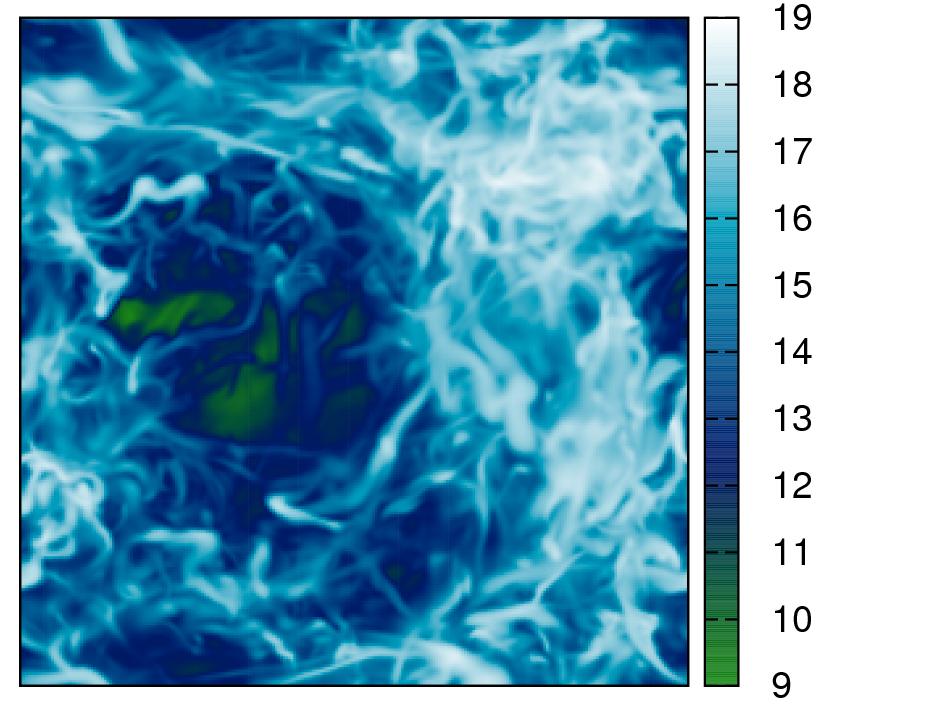}
\includegraphics[height=0.25\linewidth]{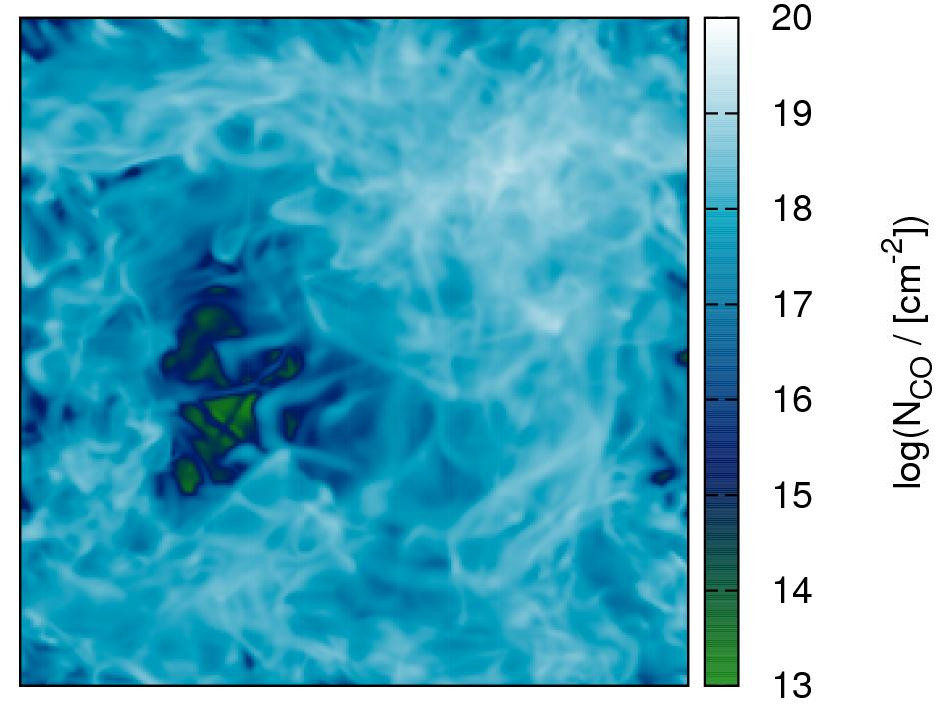}
}
\centerline{
\includegraphics[height=0.25\linewidth]{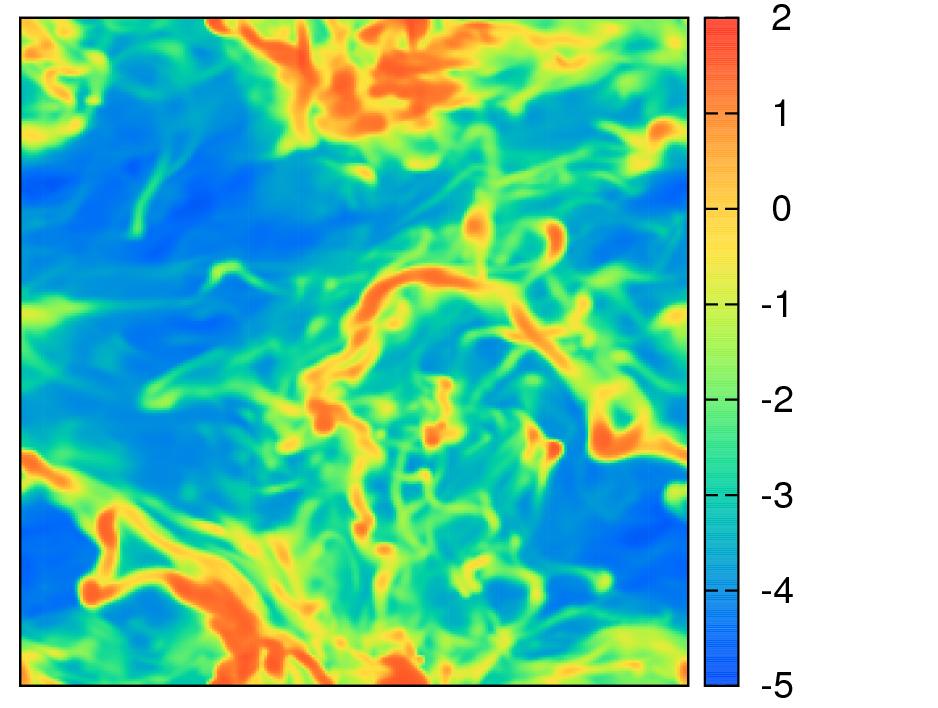}
\includegraphics[height=0.25\linewidth]{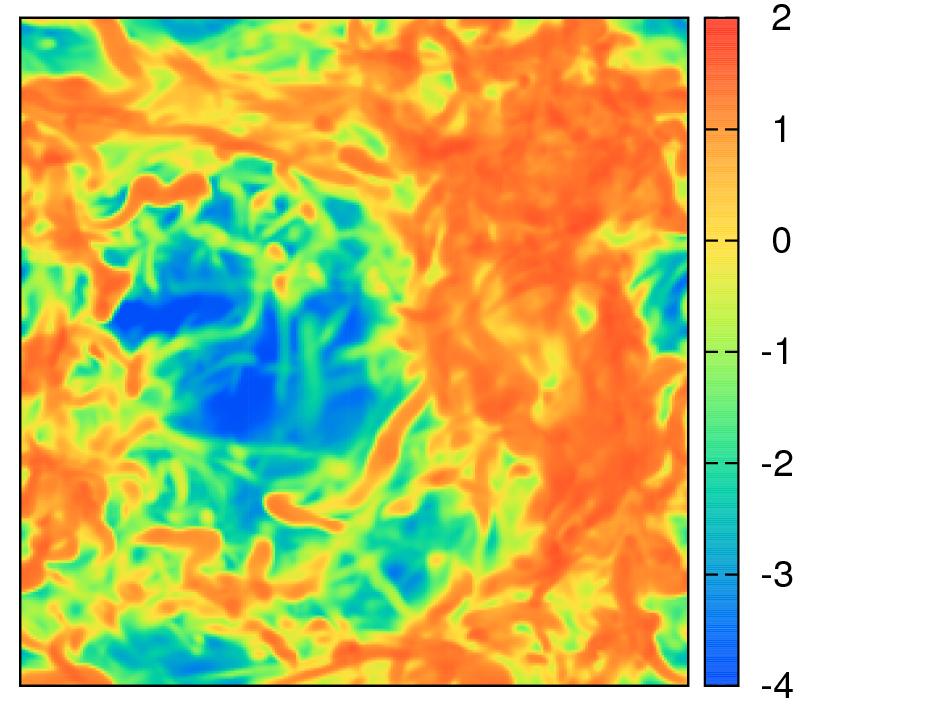}
\includegraphics[height=0.25\linewidth]{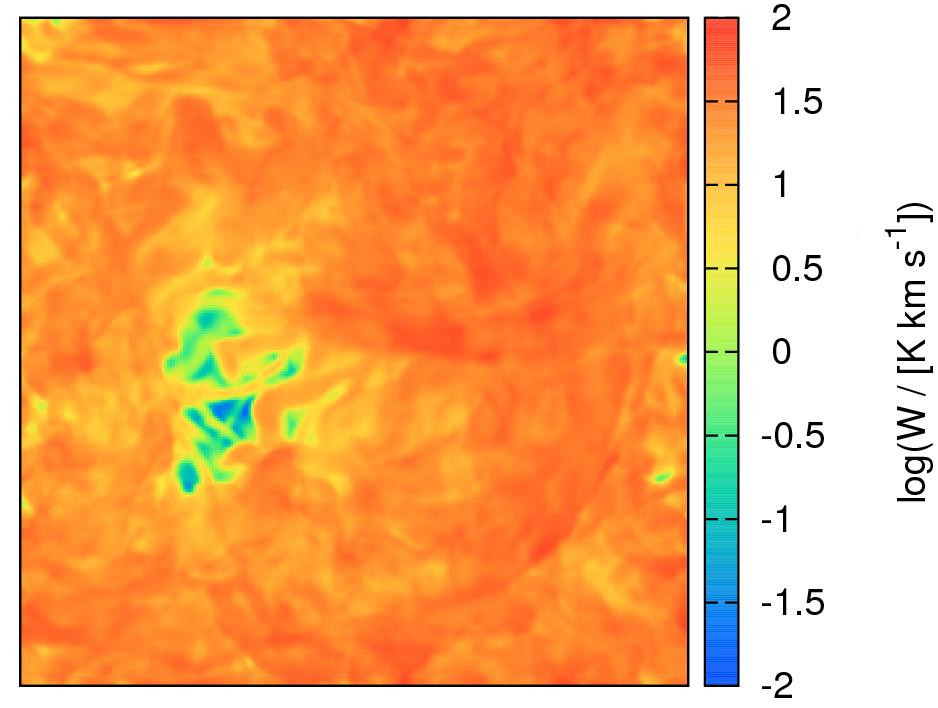}
}
\centerline{
\includegraphics[height=0.25\linewidth]{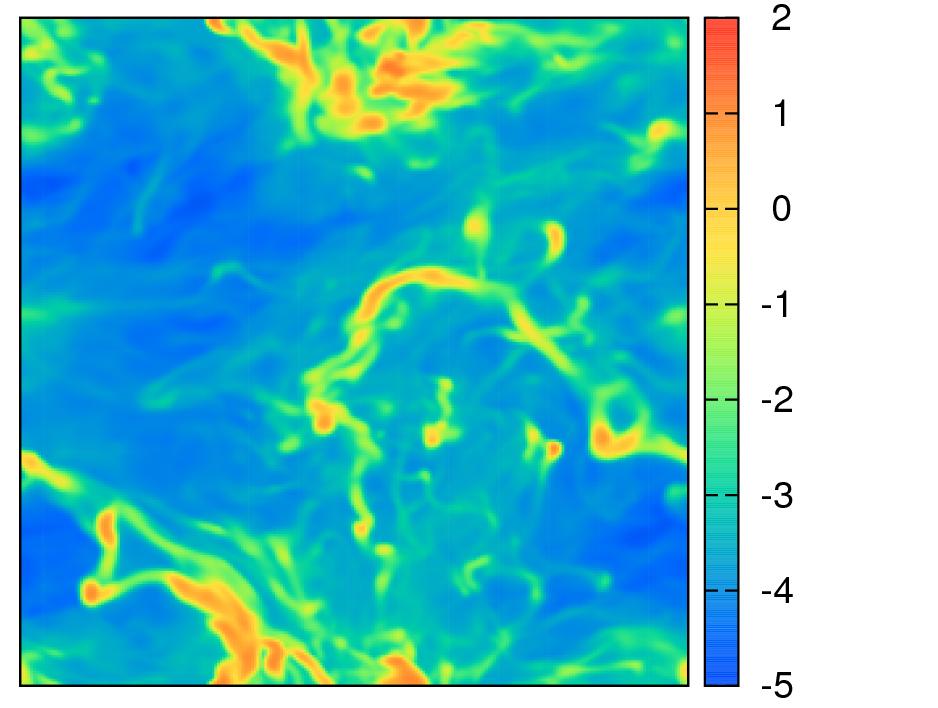}
\includegraphics[height=0.25\linewidth]{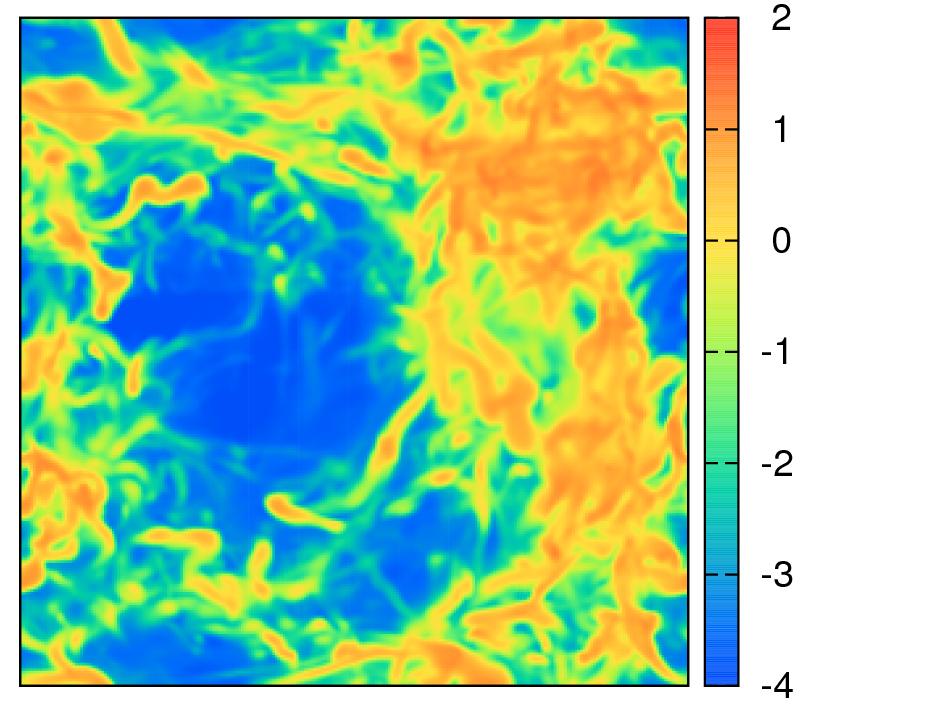}
\includegraphics[height=0.25\linewidth]{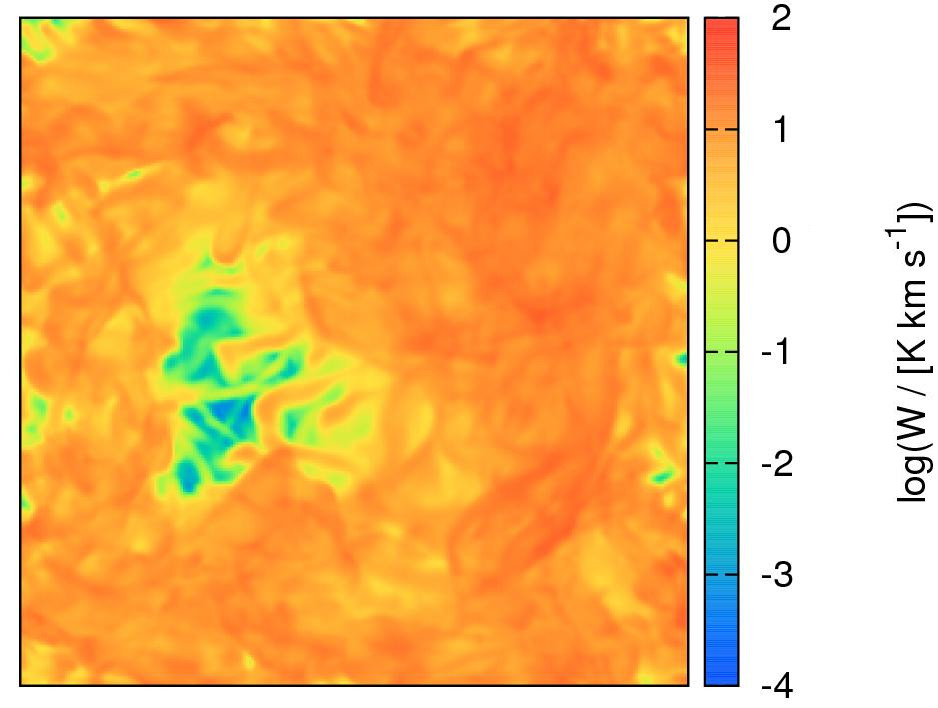}
}
\caption{Column density and integrated intensity maps computed along the LoS in z-direction for models n30 (left column), n100 (middle column) and n300 (right column). From top to bottom: column density of all gas, for H$_2$ and $^{12}$CO and integrated intensity maps of $^{12}$CO and $^{13}$CO. Note the different scaling in the colorbars. Each side has a length of $20\,$pc.}
\label{fig:images}
\end{figure*}

\begin{figure*}
\centerline{
\includegraphics[height=0.26\linewidth,width=0.42\linewidth]{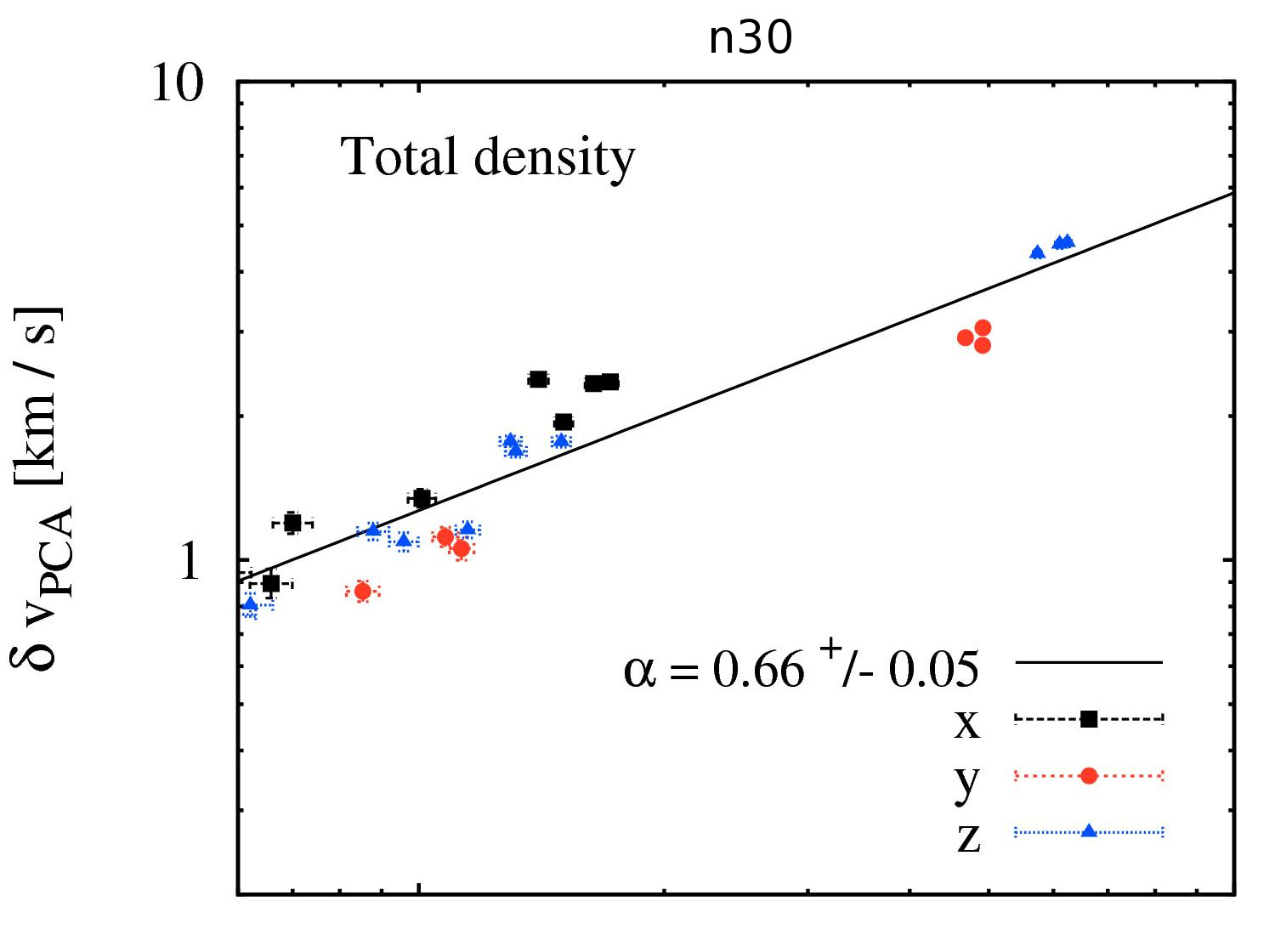}
\includegraphics[height=0.26\linewidth,width=0.35\linewidth]{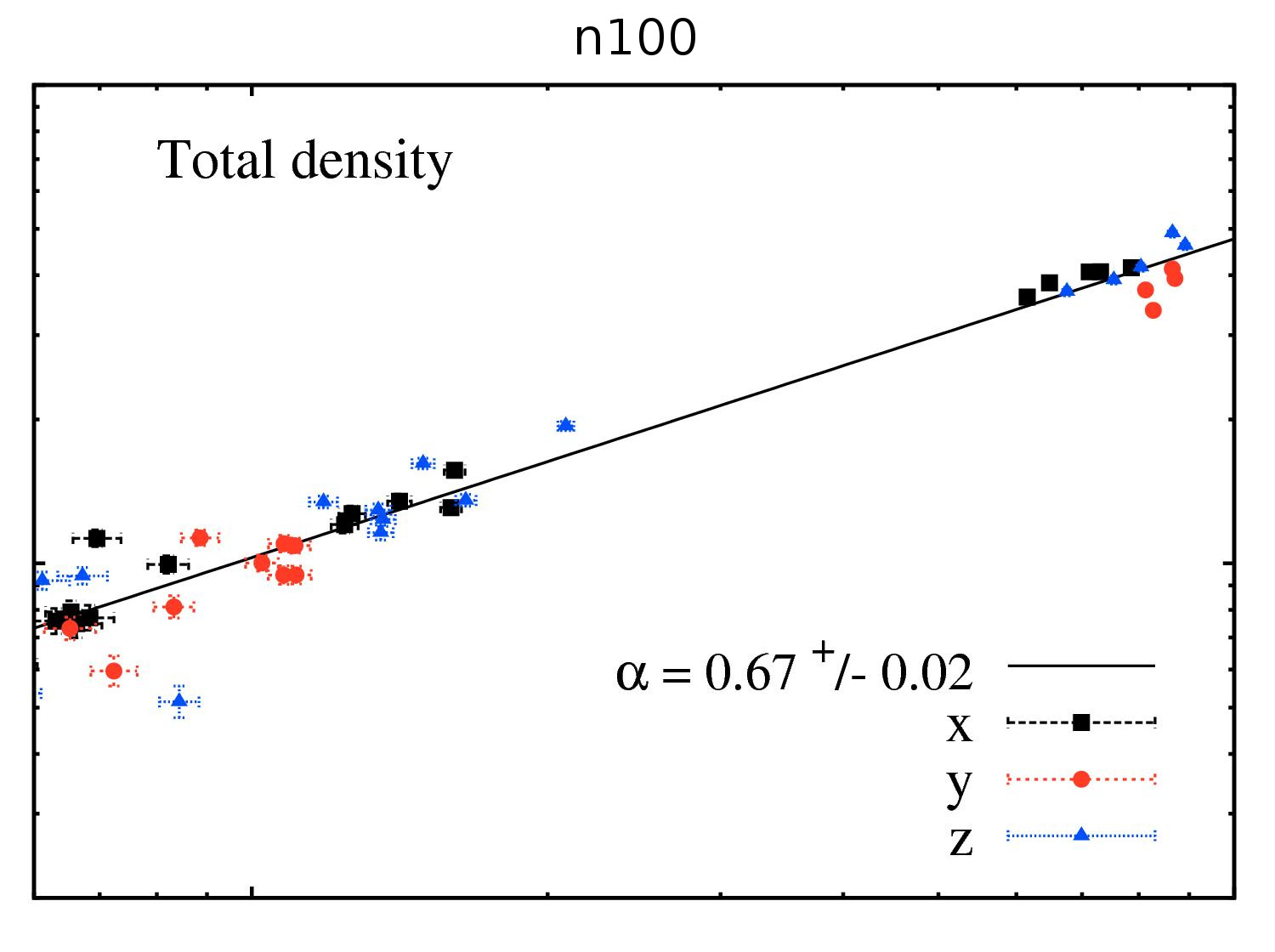}
\includegraphics[height=0.26\linewidth,width=0.342\linewidth]{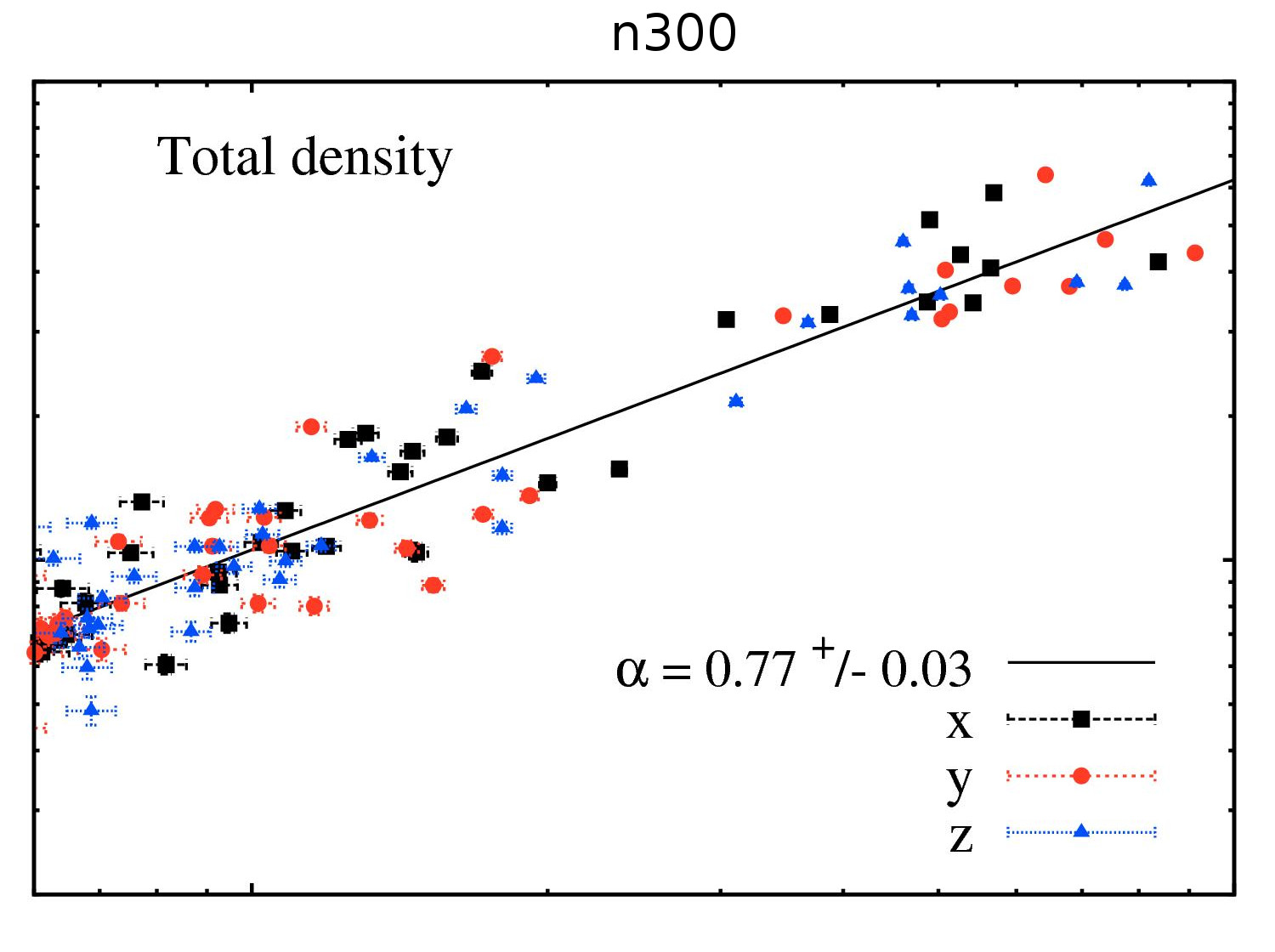}
}
\centerline{
\includegraphics[height=0.24\linewidth,width=0.42\linewidth]{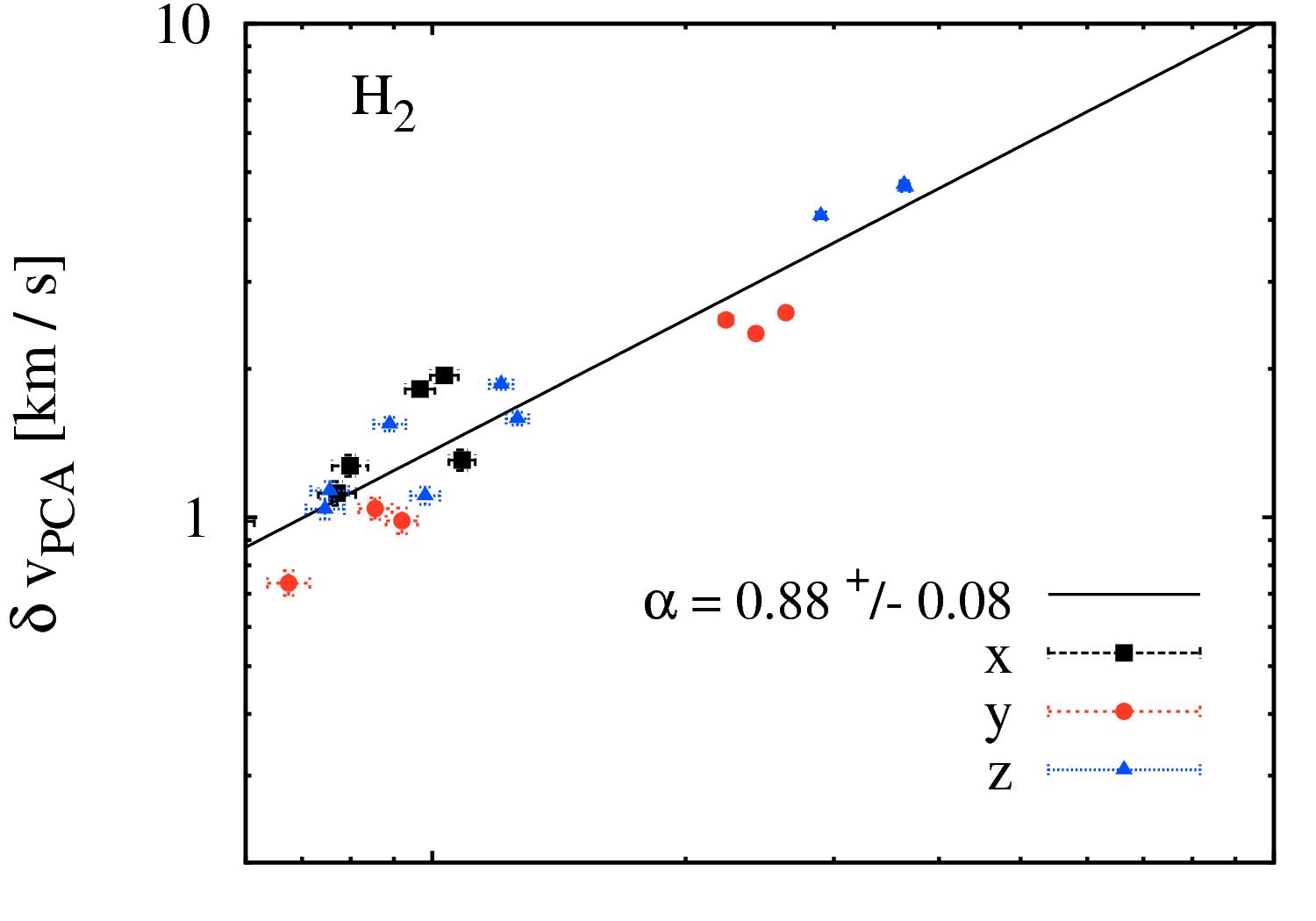}
\includegraphics[height=0.24\linewidth,width=0.35\linewidth]{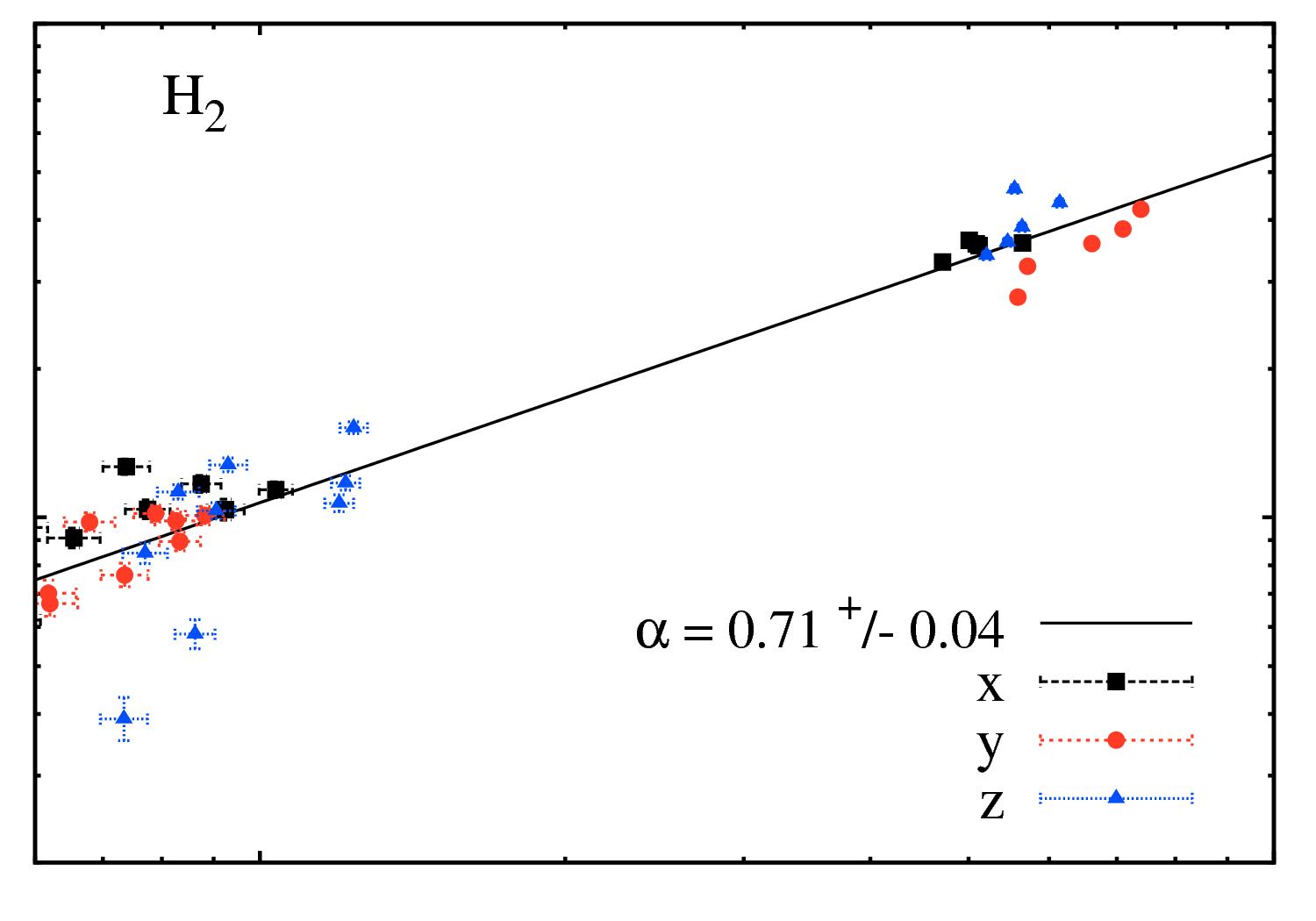}
\includegraphics[height=0.24\linewidth]{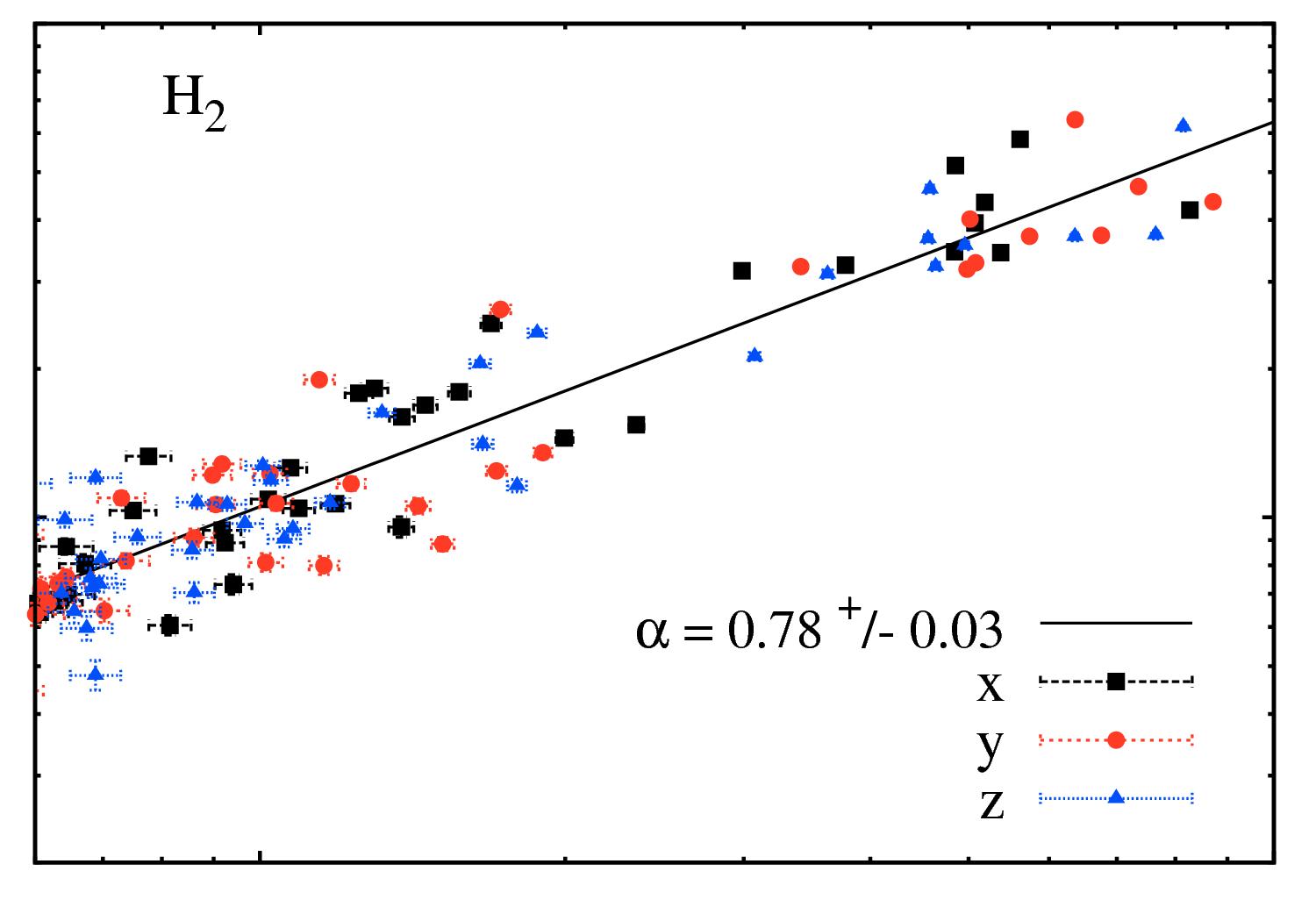}
}
\centerline{
\includegraphics[height=0.24\linewidth,width=0.42\linewidth]{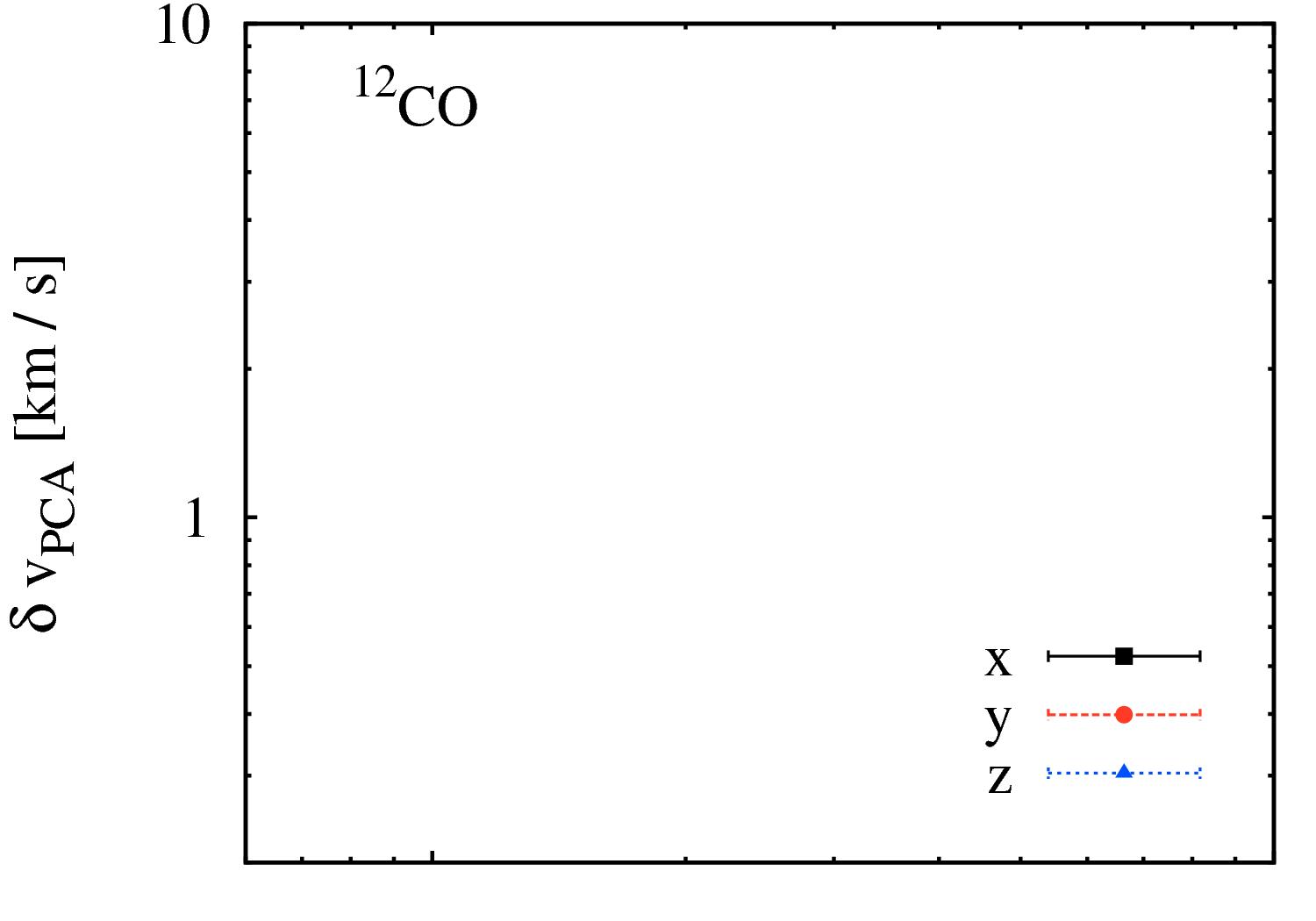}
\includegraphics[height=0.24\linewidth,width=0.35\linewidth]{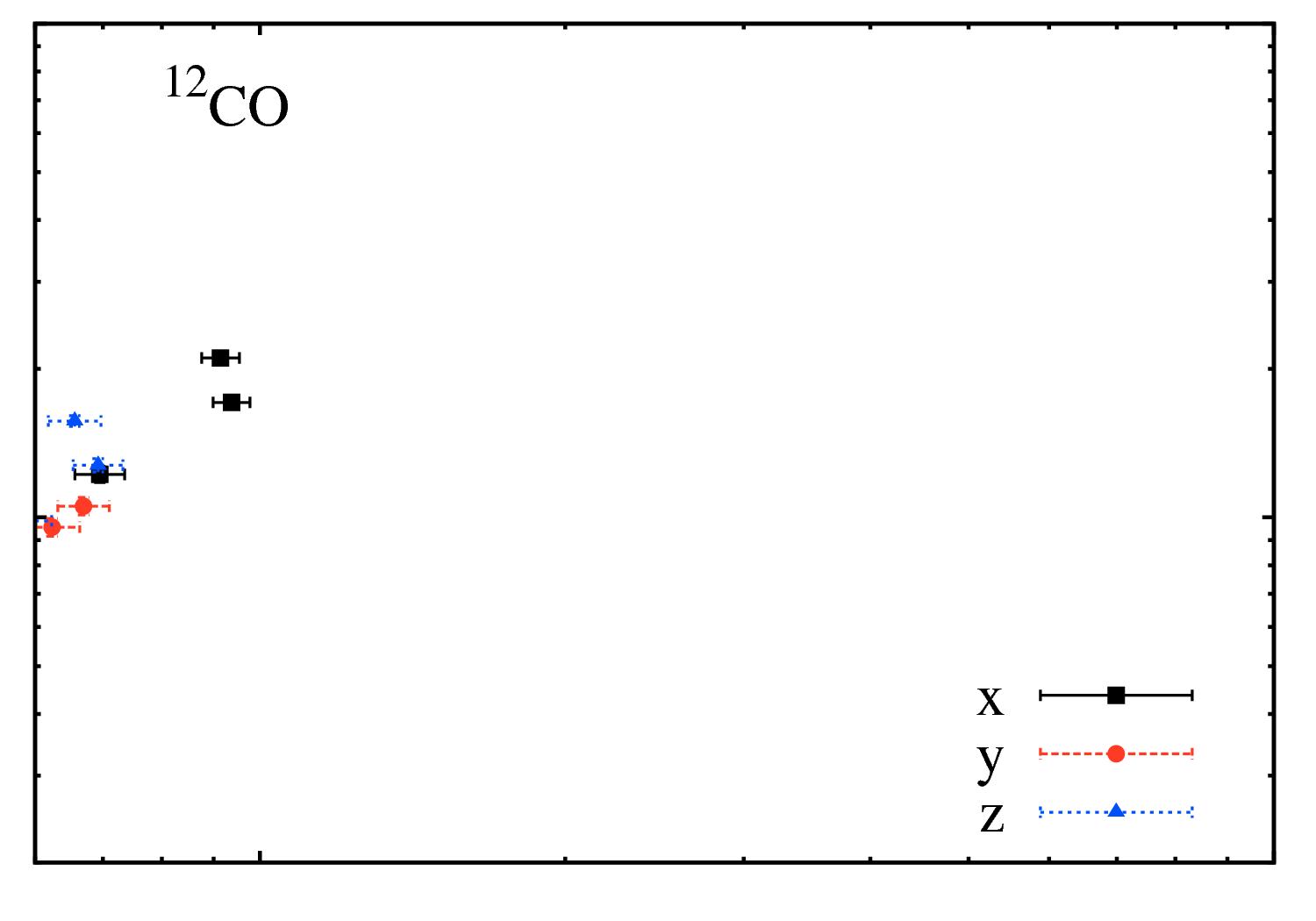}
\includegraphics[height=0.24\linewidth]{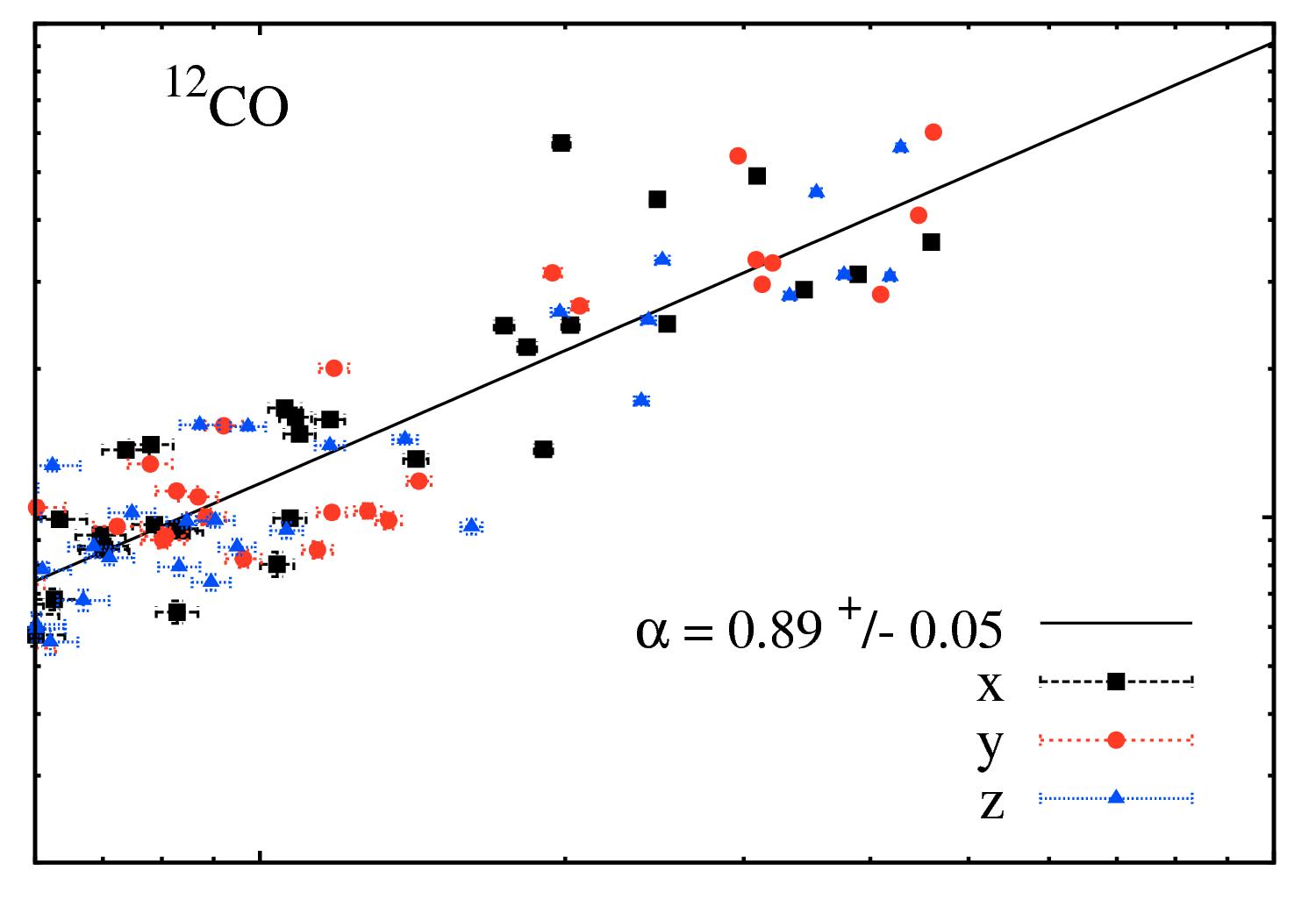}
}
\centerline{
\includegraphics[height=0.24\linewidth,width=0.42\linewidth]{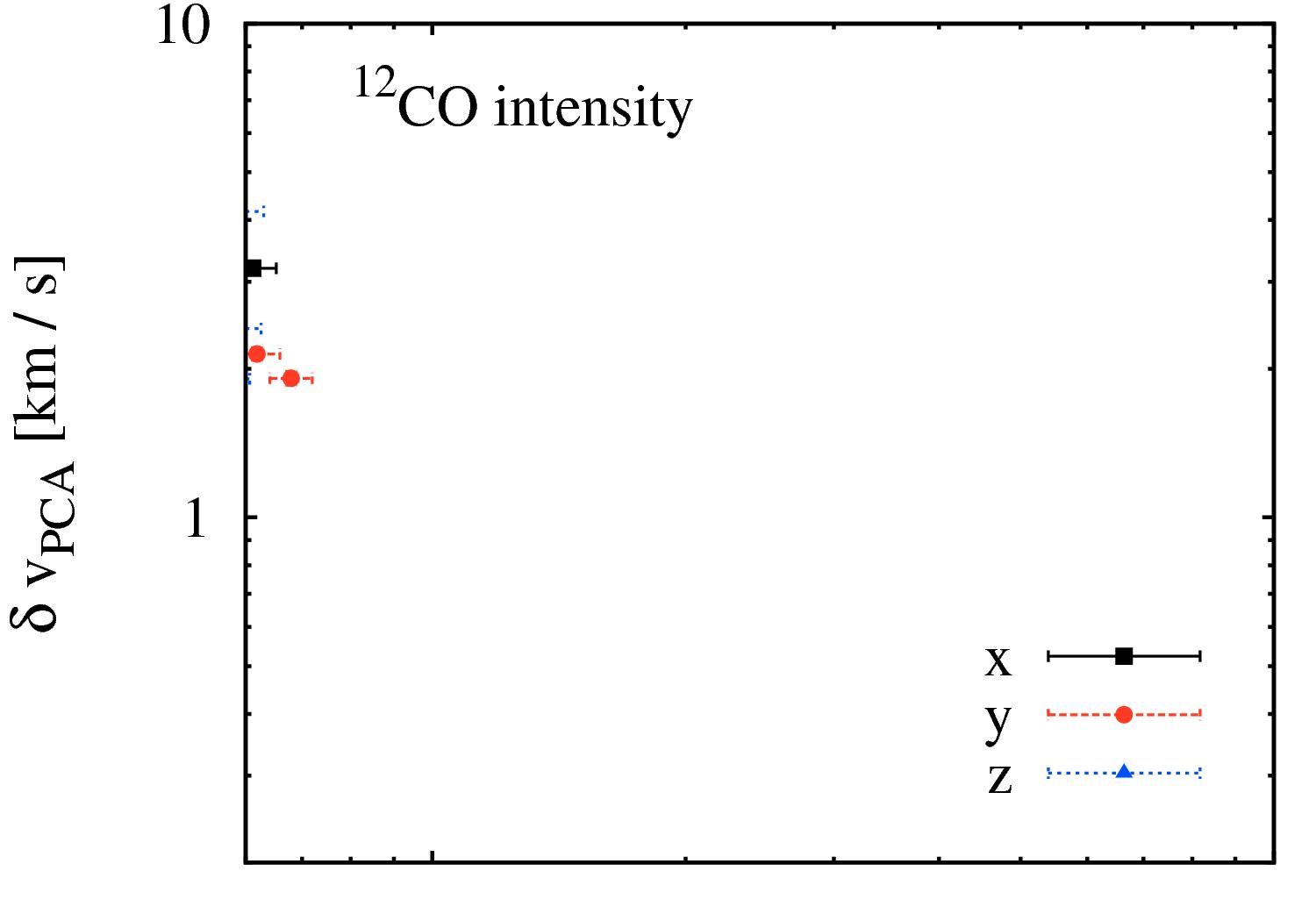}
\includegraphics[height=0.24\linewidth,width=0.35\linewidth]{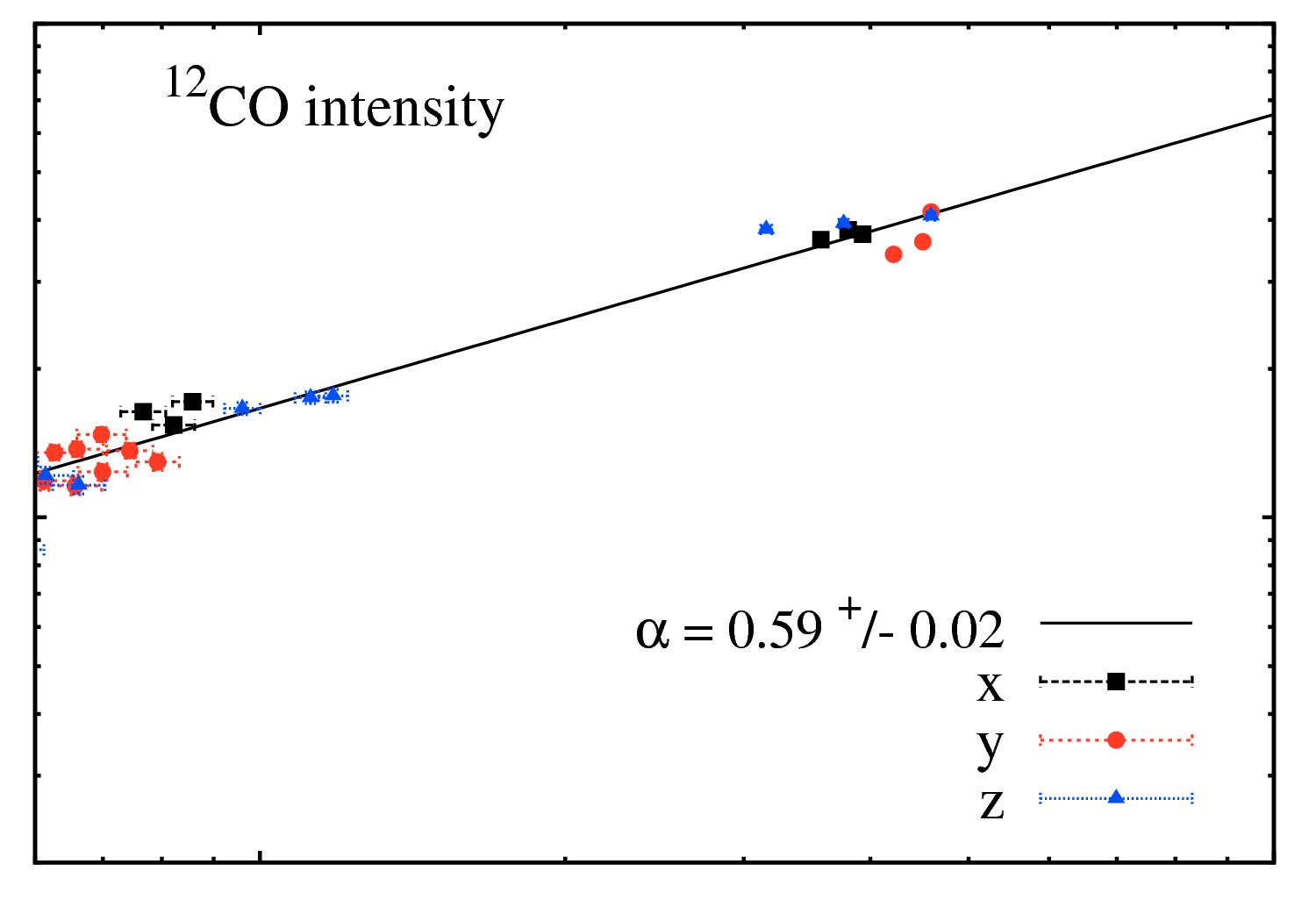}
\includegraphics[height=0.24\linewidth]{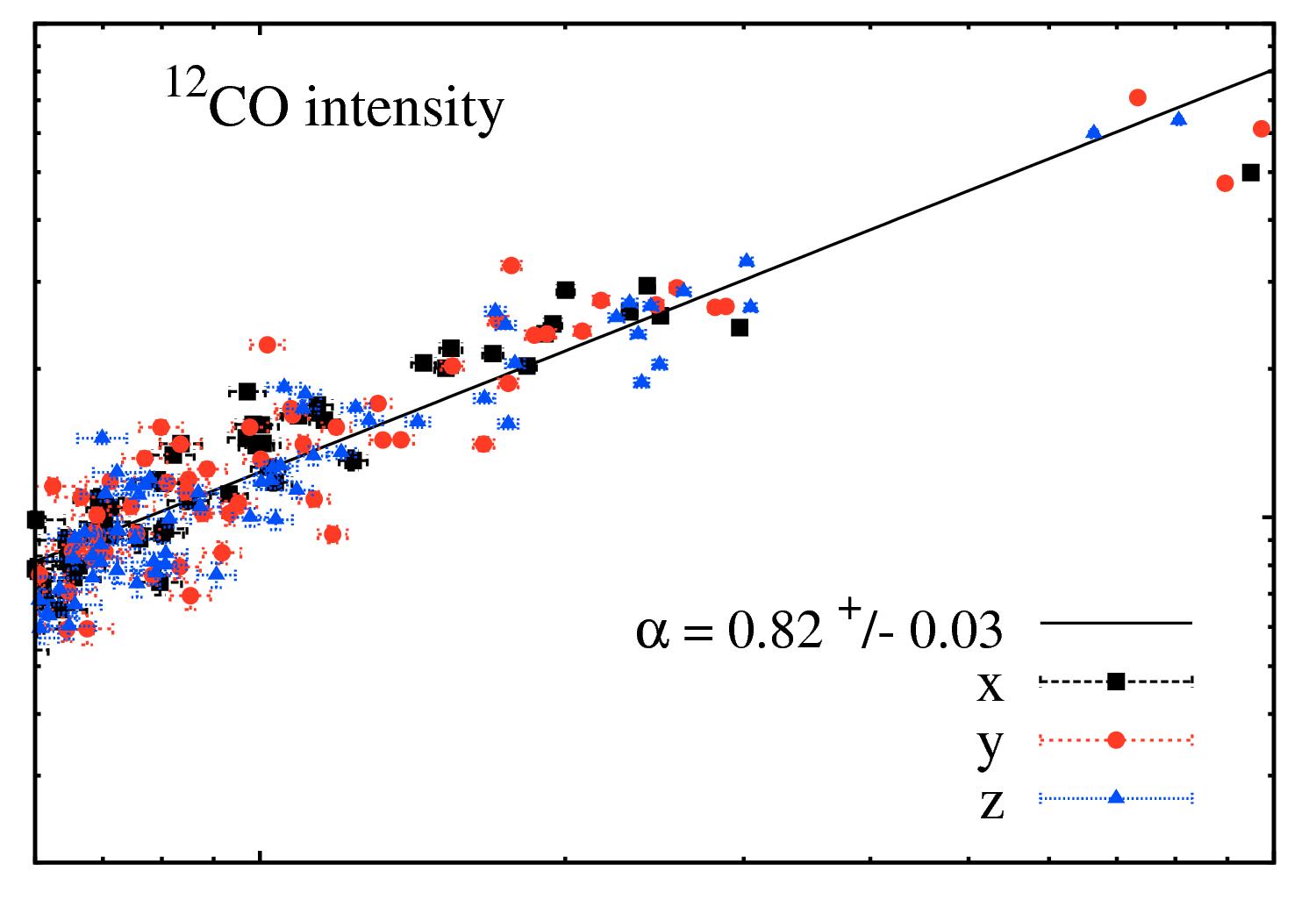}
}
\centerline{
\includegraphics[height=0.28\linewidth,width=0.42\linewidth]{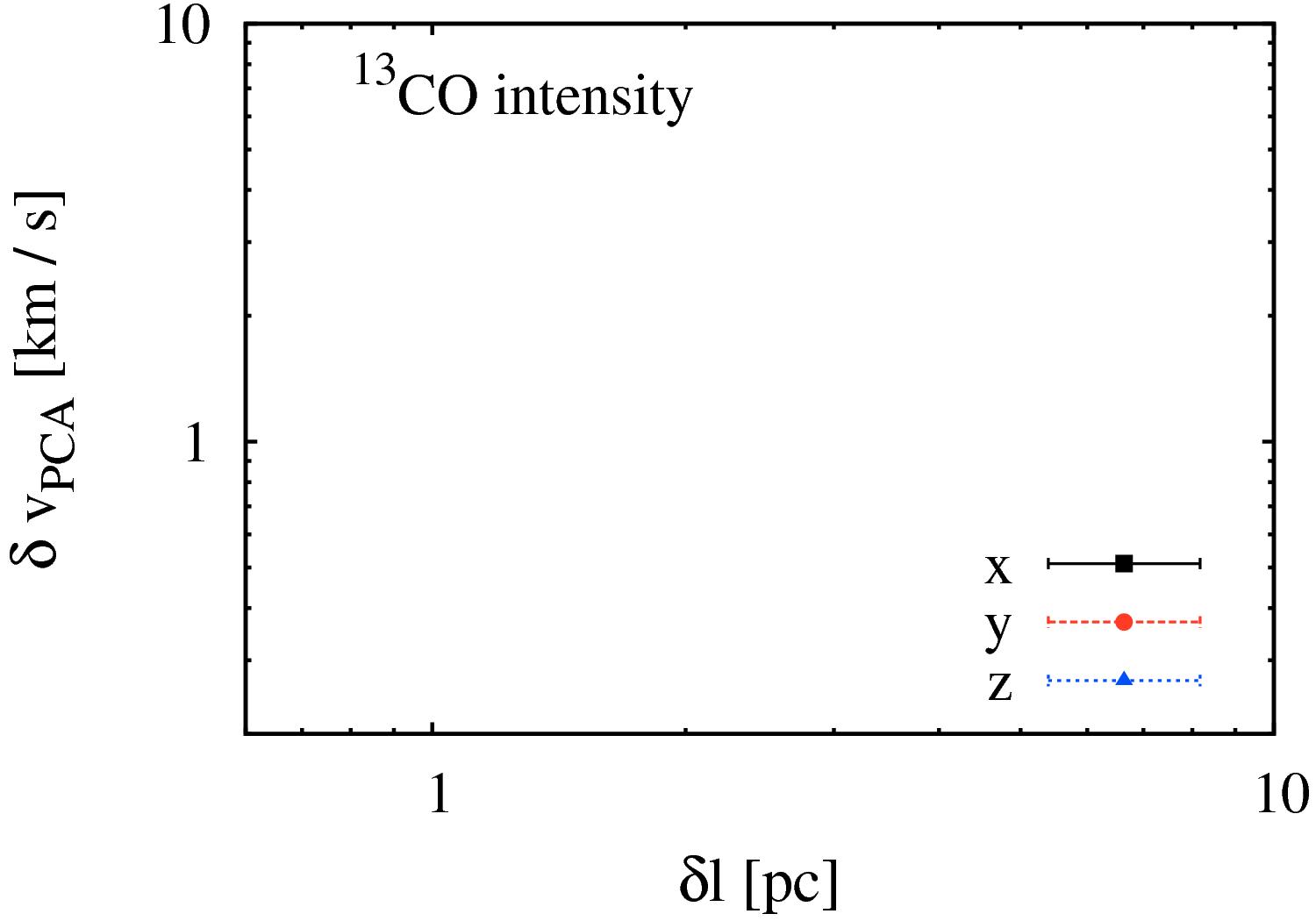}
\includegraphics[height=0.28\linewidth,width=0.35\linewidth]{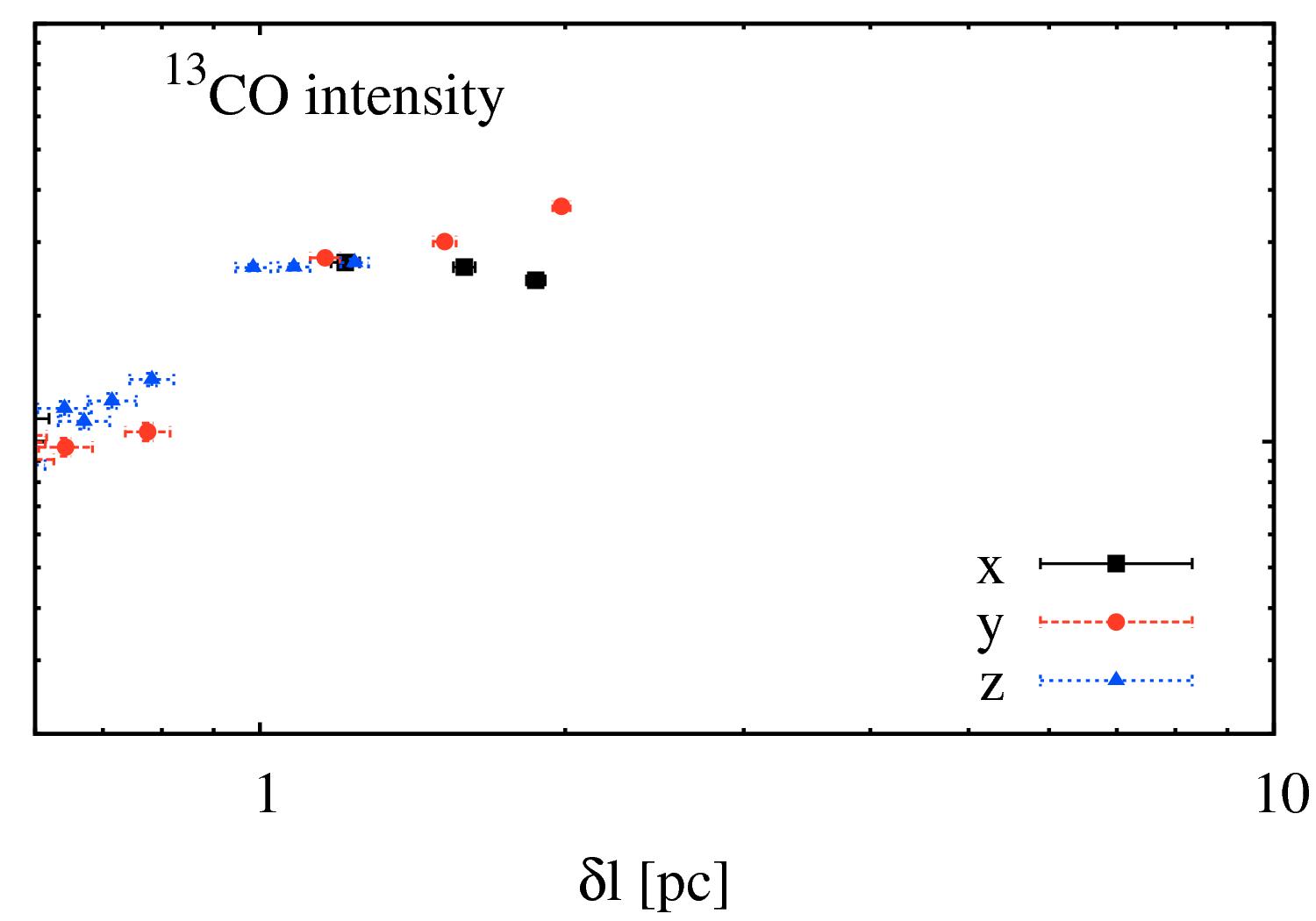}
\includegraphics[height=0.28\linewidth,width=0.343\linewidth]{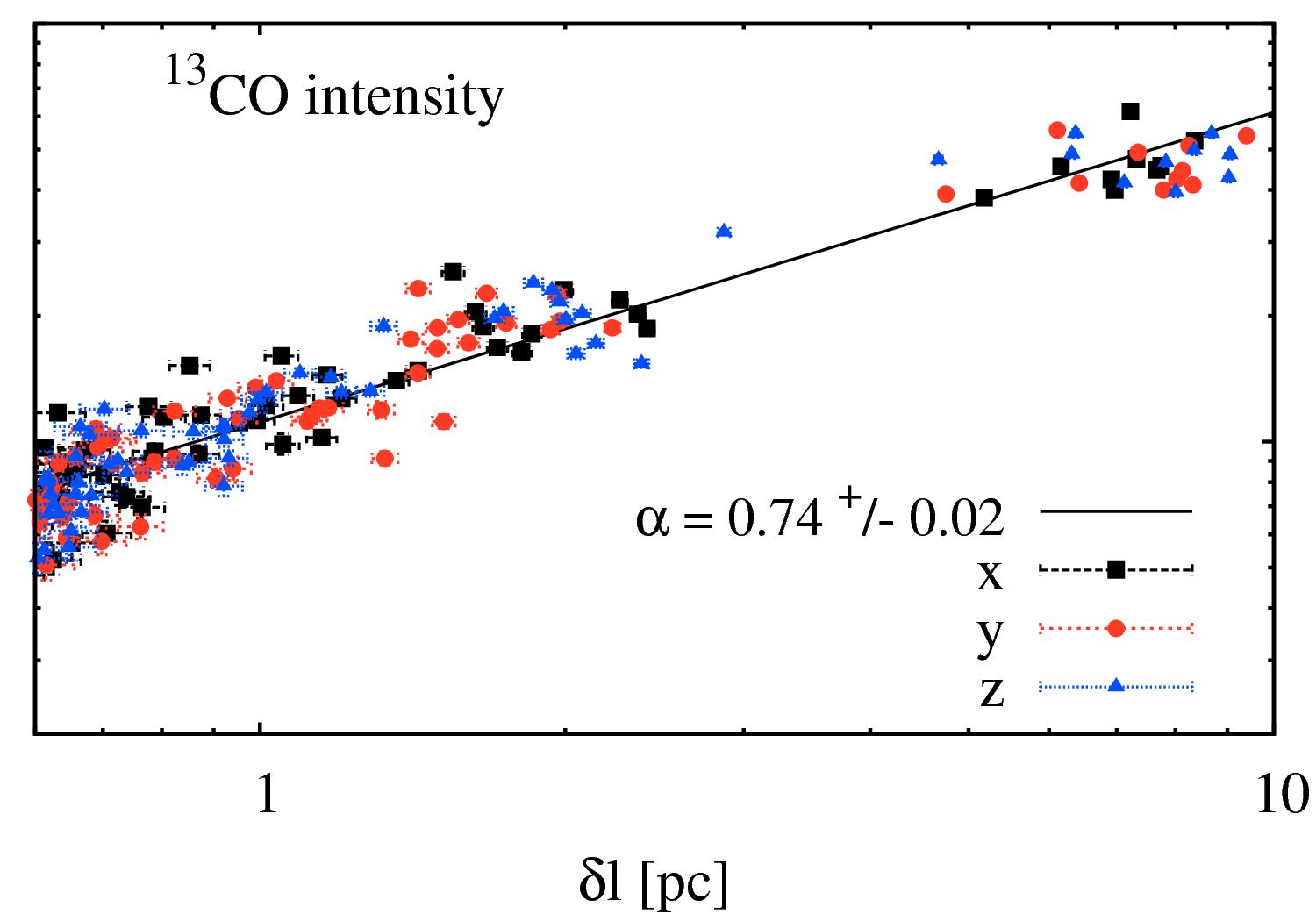}
}
\caption{Corresponding velocity fluctuations $\delta v_\text{PCA}$ as a function of spatial scales for the different components and LoS for the n30 (left column), n100 (middle column) and n300 (right column) models as presented in Figure \ref{fig:images}. Linear regression gives a PCA slope $\alpha_\text{PCA}$ (listed in each plot) for the pseudo structure function. Note that we do not find any PCA data points for the $^{12}$CO density, the $^{12}$CO integrated intensity or the $^{13}$CO integrated intensity in the n30 model, since the  abundance of CO on large scales is too low to yield large-scale PCs.}
\label{fig:PCA}
\end{figure*}

We analyse different numerical models with and without radiative transfer post-processing and vary the initial number density in the simulation box in order to study the behavior of the PCA pseudo structure function. As a reference model we choose an initial number density $n_0 = 100\,$cm$^{-3}$ for five available models, i.e. for the total number density, H$_2$ and $^{12}$CO number density and for the intensities of the $J = 1 \rightarrow 0$ line of $^{12}$CO and $^{13}$CO. We chose this value of $n_{0}$ for our reference model because it reflects the typical density of molecular clouds in the solar neighbourhood accessible by observations. We also analyse models with initial number densities $n_0 = 30\,$cm$^{-3}$ and $n_0 = 300\,$cm$^{-3}$ to span a range of possible densities in common MCs \citep{RomanDuvalEtAl2010}. In Table~\ref{tab:setup}, we give an overview of the numerical models that we analyse in this paper and list a number of values characterising the chemical and thermal state of the gas, such as the mean H$_{2}$ and CO number densities and column densities, and the mean temperature of the gas. We quote both mass- and volume-weighted quantities, which we define using the expressions $\langle f \rangle_{mass} = \sum f \rho dV / \sum \rho dV$ and $\langle f \rangle_{vol} = \sum f dV / \sum dV$, respectively. We also quote the mean (area-averaged) integrated intensities of $^{12}$CO and $^{13}$CO. More detailed analyses of the H$_{2}$ and CO distributions produced in this kind of turbulent simulation and the resulting distributions of $^{12}$CO integrated intensities can be found in the works by \citet{GloverEtAl2010} and \citeauthor{ShettyEtAl2011a}~(2011a); in the interests of brevity, we do not repeat their analysis here. Integerated intensity and column density PDFs are shown in \citet{GloverEtAl2010} and \citeauthor{ShettyEtAl2011a}~(2011a). Figure \ref{fig:images} shows column density and integrated intensity maps of all models and for all components. Figure \ref{fig:PCA} shows the corresponding results from the PCA including the slopes of the pseudo structure function which are summarized in Table \ref{tab:PCAvalues}.

\begin{table*}
\begin{tabular}{c|c|c|c|c|c}
\hline\hline
 Model name & Total density & H$_2$ & $^{12}$CO & $^{12}$CO intensity & $^{13}$CO intensity \\
 & $\alpha_\text{PCA}$ & $\alpha_\text{PCA}$ & $\alpha_\text{PCA}$ & $\alpha_\text{PCA}$ & $\alpha_\text{PCA}$ \\
\hline
 n30 & $0.66 \pm 0.05$ & $0.88 \pm 0.08$ & --- & --- & --- \\
 n100 & $0.67 \pm 0.02$ & $0.71 \pm 0.04$ & --- & $0.59 \pm 0.02$ & --- \\
 n300 & $0.77 \pm 0.03$ & $0.78 \pm 0.03$ & $0.89 \pm 0.05$ & $0.82 \pm 0.03$ & $0.74 \pm 0.02$ \\
\hline\hline
\end{tabular}
\caption{Slopes $\alpha_\text{PCA}$ of the pseudo structure function for our different models described in Table \ref{tab:setup}.}
\label{tab:PCAvalues}
\end{table*}

\subsection{Analysis of the n100 reference model}
\label{subsec:PCAn100}

The middle column of Figure \ref{fig:images} shows maps of all components of our fiducial n100 model. The total and H$_2$ density are similar, since most of the gas consists of H$_2$: around 70\% of the hydrogen is molecular by the end of the simulation. The CO column density map also traces the same structures, but spans a much wider range of values, demonstrating that CO has very low abundance along the low column density sight-lines. Finally, the intensity maps largely reflect the CO distribution, although the fact that the $^{12}$CO emission is optically thick in places means that some of the highest density peaks in the density maps do not correspond to the brightest regions in the $^{12}$CO intensity map (see also \citeauthor{ShettyEtAl2011a}~2011a).

Detailed comparison of the $^{12}$CO intensity map with the $^{12}$CO density map shows that the intensity map is smoother and spans a smaller range of values (6 orders of magnitude in intensity compared to 10 orders of magnitude in density). This occurs because the $J=1 \rightarrow 0$ line of $^{12}$CO is easily excited and can be bright even in low-density regions. For $^{13}$CO the abundance is lower by a factor of 50 compared to $^{12}$CO, but since it is optically thin, the peaks coincide with those of the CO density.

The middle column of Figure \ref{fig:PCA} shows the ``composite PCA pseudo structure function'', including data from 3 LoS directions and 5 time snapshots. We obtain slopes of the pseudo structure function via linear regression in log-log space using data of five snapshots for each chemical species. In general, we find an increasing trend of $\delta v_\text{PCA}$ with increasing spatial scales. As shown in \citet{GloverEtAl2010}, the simulations have reached a steady state in both the dynamical and chemical properties.

For the H$_2$ and the total number density PPV cubes, the slope of the PCA pseudo structure function is found to be $0.71 \pm 0.04$ and $0.67 \pm 0.02$, respectively. Both exponents are similar, because most of the gas is made up of H$_2$, as is evident in Figure \ref{fig:images}. The $^{12}$CO intensity yields a PCA slope of $\alpha_\text{PCA} = 0.59 \pm 0.02$. We do not fit the $^{13}$CO intensity and $^{12}$CO density (see Figure \ref{fig:PCA}) since PCA does not find structures on larger scales necessary to perform a robust fit.

For the PPV cubes resulting from the total density, H$_2$ and $^{12}$CO intensity we find a distinct gap between \mbox{$\sim$2-$6\,$pc} for the total density PPV cube, \mbox{$\sim$1.2-$5\,$pc} for the H$_2$ density PPV cube and \mbox{$\sim$1.2-$3\,$pc} for the $^{12}$CO intensity PPV cube. This gap is made up by the width and the symmetry of the 0th and 1st order eigenvectors. Furthermore, we analyse five snapshots of a simulation over $\Delta t_\text{snap} = 0.32\,$Myr. This time interval is short compared to the shock crossing time $\Delta t_\text{cross} = 1.90\,$Myr, hence the overall morphology changes only slightly. If the 0th eigenvector is axially symmetric and has a large width we mainly observe velocity fluctuations on larger scales since the eigenvector is projected onto the PPV cube. In numerous cases we find that the 1st order eigenvector has about half of the width of the 0th eigenvector and the 2nd component about half of the width of the 1st component and so on, i.e. every incremental eigenvector has a smaller width. Thus, the gap that we find in the data mostly lies between the 0th and 1st order eigenvectors. Since we do not find any data points in the $^{12}$CO density above 1$\,$pc indicates that PCA cannot detect any coherent structures at the largest scales in this particular case.

\citet{BruntAndHeyer2013} investigated how to use PCA as a statistical tool to analyse the structure and dynamics of the ISM, with emphasis on the study of turbulence. If the structures in PPV space are very fragmented, PCA can operate more as a cloud finder than a descriptor of kinematic turbulence, as shown by \citet{HeyerAndSchloerb1997}. Although there might be some discrete features along the LoS in PPV space in our simulations, the overall structure and dynamics are turbulent such that PCA describes the kinematics of the physical fields very well.

\subsection{Dependence of the PCA results on initial density}
\label{subsec:parameter}

\begin{table*}
\begin{tabular}{c||c|c|c|c|c}
\hline\hline
 Model name & Total density & H$_2$ & $^{12}$CO & $^{12}$CO intensity & $^{13}$CO intensity \\
\hline
 n30 & $0.32 \pm 0.15$ & $0.40 \pm 0.12$ & --- & --- & --- \\
 n100 & $0.18 \pm 0.04$ & $0.16 \pm 0.04$ & --- & $0.21 \pm 0.07$ & --- \\
 n300 & $0.28 \pm 0.15$ & $0.28 \pm 0.15$ & $0.36 \pm 0.18$ & $0.52 \pm 0.21$ & $0.30 \pm 0.13$ \\
\hline\hline
\end{tabular}
\caption{Mean and standard deviation of $l_1 / l_0$ for all density models averaged over all snapshots and LoS. We find mean ratios of $\langle l_1 / l_0 \rangle \gtrsim 0.2$ for all models and components. Note that the numbering of scales in \citet{BruntEtAl2009} starts with 1 and not with 0 as in this paper.}
\label{tab:l2l1ratio}
\end{table*}

\begin{figure}
\centerline{
\includegraphics[width=1.0\linewidth]{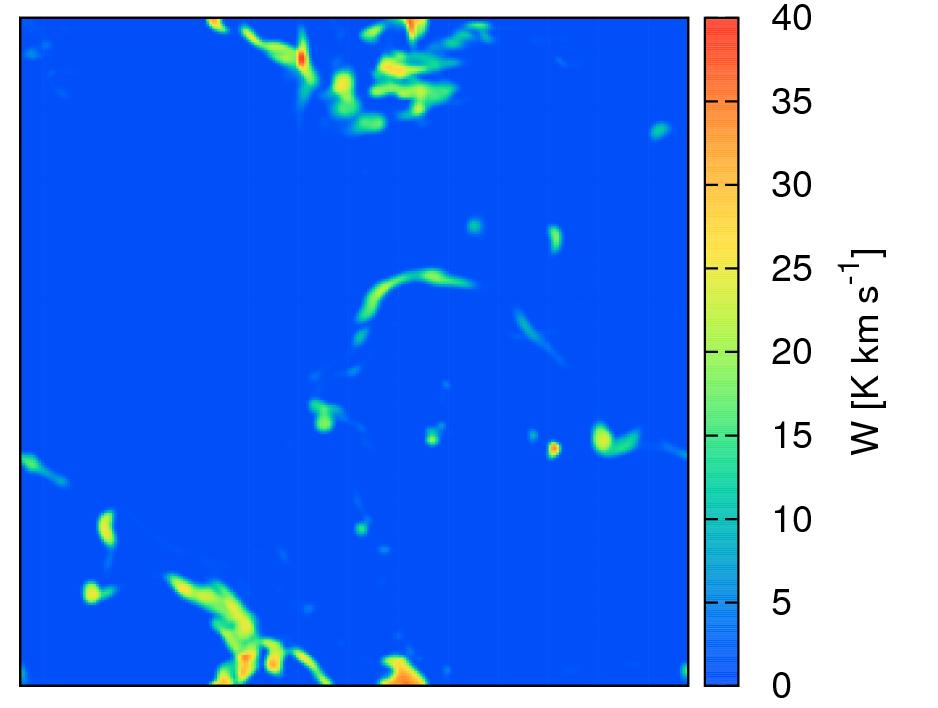} \\
}
\centerline{
\includegraphics[width=1.0\linewidth]{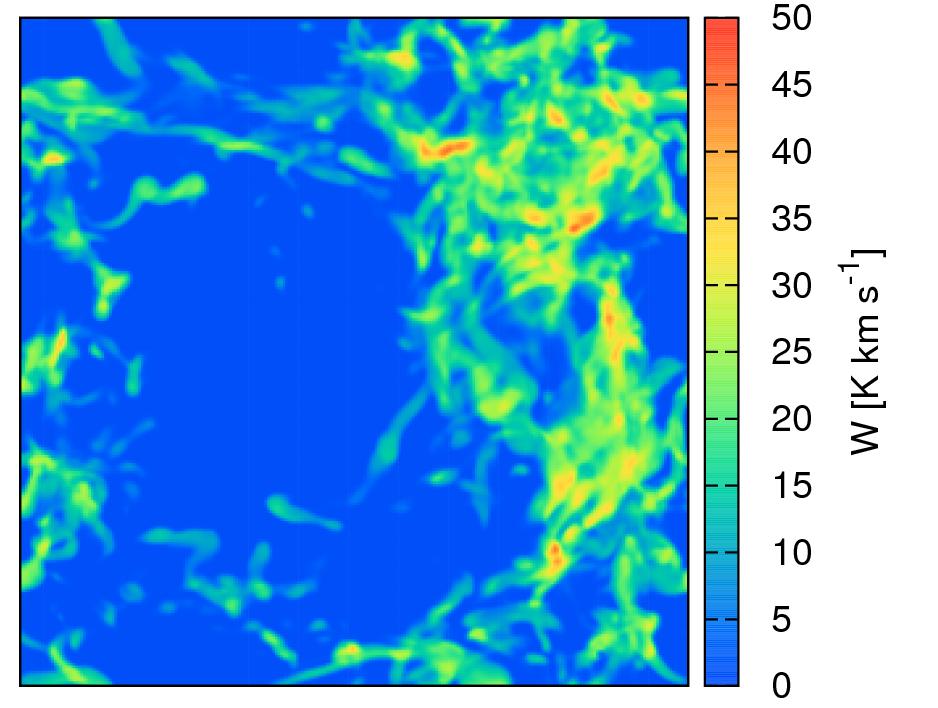} \\
}
\centerline{
\includegraphics[width=1.0\linewidth]{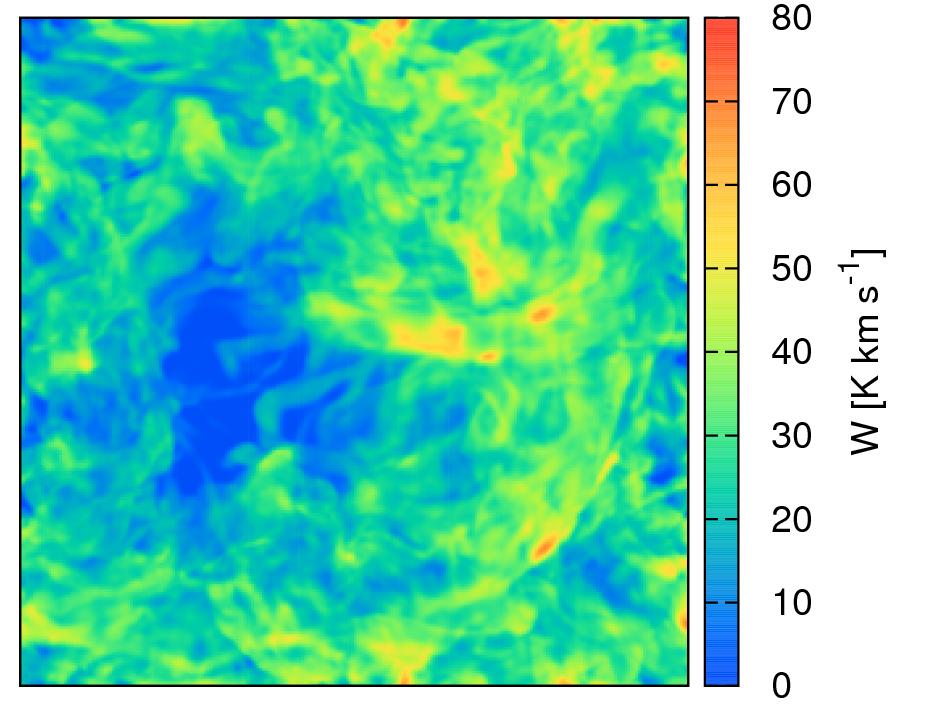} \\
}
\caption{Integrated intensity maps for $^{12}$CO on a linear scale computed along the LoS in z-direction for different times and turbulent velocity fields for models n30 (top), n100 (middle) and n300 (bottom). In model n30 there is not enough CO emission produced at low densities to yield a large scale PC in the analysis. Note the different scaling in the colorbars.}
\label{fig:linearscale}
\end{figure}

In order to investigate how the density affects the result of the PCA, we analyse two models, n30 and n300 (Figure \ref{fig:images} and Table \ref{tab:PCAvalues}) in addition to our fiducial n100 model. Our analysis is identical to the one described above.

For the n300 model (see Appendix \ref{app:realizations}) linear regression yields slightly higher values for the total number density and H$_2$ density PPV cubes compared to the n100 model. We find slopes of $0.78 \pm 0.03$ and $0.77 \pm 0.03$, which are again statistically identical since 94\% of the initial atomic hydrogen is converted into H$_2$ (see Table \ref{tab:setup}). For the $^{12}$CO density we find a value of $0.89 \pm 0.05$ and for the $^{12}$CO intensity we obtain a value of $0.82 \pm 0.03$. The $^{13}$CO intensity yields a value of $0.74 \pm 0.02$.

For the n30 model we use three available snapshots to compute individual PCA data points for each LoS direction. We find different slopes of $0.66 \pm 0.05$ and $0.88 \pm 0.08$ for the total number density and for H$_2$ since only $\approx\,$35\% of the atomic hydrogen is converted into H$_2$ for the n30 model (see Table \ref{tab:setup}). For the $^{12}$CO and $^{13}$CO intensity there is not enough CO emission produced at low densities to yield large-scale PCs in the analysis (see Figure \ref{fig:PCA}). The reason that the method fails in this case can be seen quite clearly in Figure \ref{fig:linearscale}, where we show maps of $^{12}$CO integrated intensity on a linear scale for all three models. In the low density models, the CO distribution is dominated by just a few small, bright peaks, and there is very little emission from the intervening diffuse gas. The CO distribution is therefore far too intermittent to yield a meaningful slope for the pseudo structure function in this case.

The differences in the PCA pseudo structure functions between the different density runs, the different chemical species, the density and intensity results are discussed in Section \ref{sec:discussion}.

\subsection{Effects of the turbulence driving scale}
\label{subsec:drivingscale}

We also compare our analysis with studies by \citet{OssenkopfAndMacLow2002} and \citet{BruntEtAl2009} who investigate the correlation of the ratio $l_2 / l_1$ in both numerical simulations and observations with the driving scale of turbulence. They find that only models driven at large scales are consistent with observations and obtain mainly values of $l_2 / l_1 \gtrsim 0.2$ from $^{12}$CO observations of real MCs \citep[see, e.g., Figure 5 in][]{BruntEtAl2009}. Table \ref{tab:l2l1ratio} shows that we find mean ratios of $\langle l_1 / l_0 \rangle \gtrsim 0.2$ for all models and components. Note that the numbering of scales in \citet{BruntEtAl2009} starts with 1 and not with 0 as in this paper. As we use a large scale driving at modes of $k = 1-2$ for all models we can confirm the results of \citet{OssenkopfAndMacLow2002} and \citet{BruntEtAl2009}, i.e. the correlation between the mean ratios of $l_1 / l_0$ with the driving scale, which is not significantly affected by the chemistry and the radiative transfer in our simulations.

\section{Discussion}
\label{sec:discussion}

\subsection{Comparison of theoretical and observational results}
\label{comparisonofn300}

\begin{figure}
\centerline{
\includegraphics[width=1.0\linewidth]{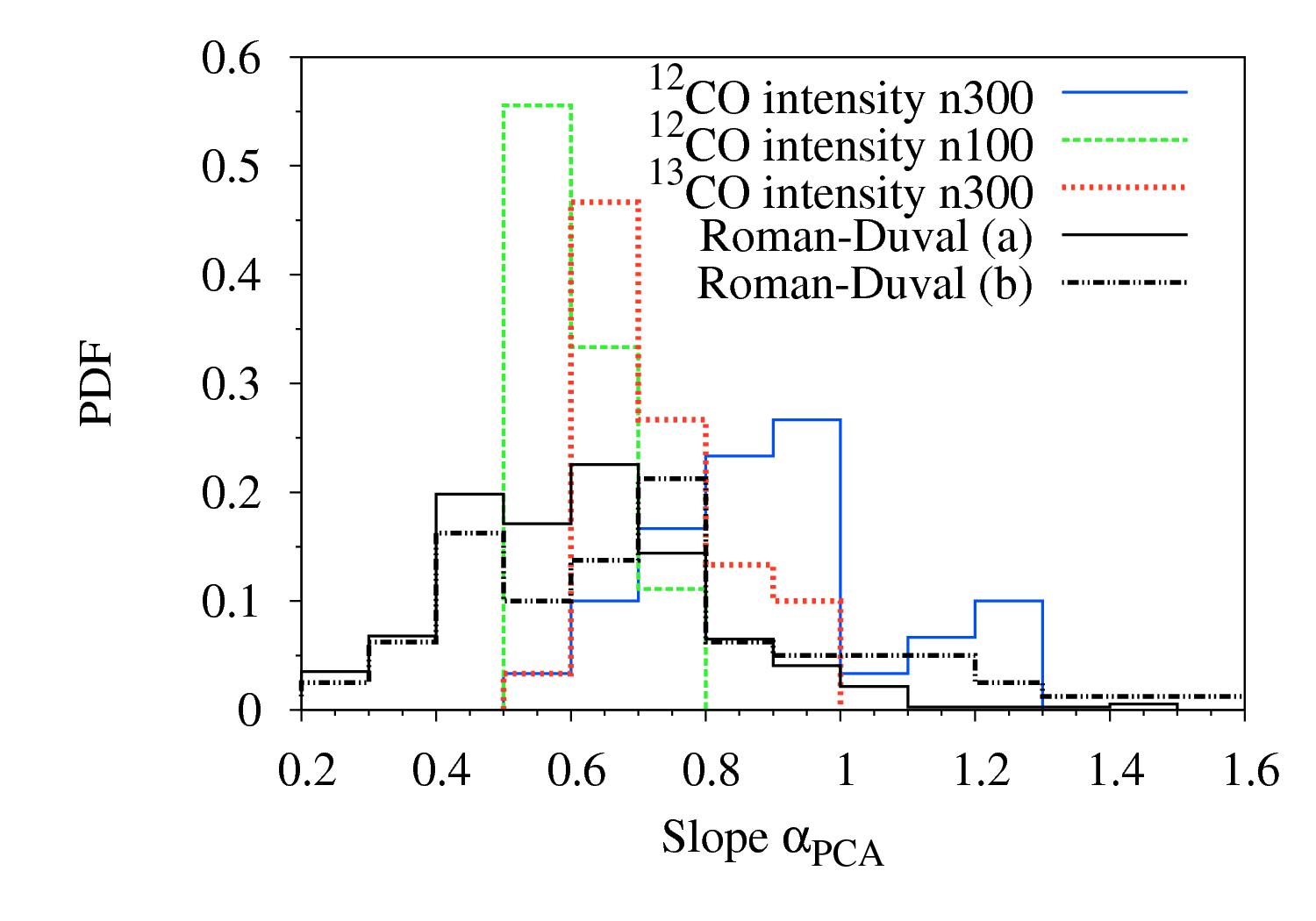}
}
\caption{Probability density functions (PDFs) for the slope of the pseudo structure function derived from our n100 and n300 models. These PDFs quantify the variance between different time snapshots and different lines of sight though our simulation volumes. We show results for both the $^{12}$CO and the $^{13}$CO intensity. For comparison, we also show the values derived by \citet{RomanDuvalEtAl2011} from their Galactic Ring Survey $^{13}$CO observations (see, in particular, their Figure~10). The values denoted as Roman-Duval (a) were computed using a sample that included all molecular clouds with angular sizes larger than 48'' (corresponding to the FWHM of the observations), while values denoted as Roman-Duval (b) were computed using a smaller sample that included only those clouds with angular sizes greater than 96'' (i.e.\ twice the FWHM).} 
\label{fig:hist}
\end{figure}

We now compare our PCA results to those obtained from observations of Galactic molecular clouds. Figure \ref{fig:hist} shows a PDF of the PCA slopes for all snapshots and LoS directions for the n100 and n300 models as described in Section \ref{sec:results}, together with the inferred slopes from Figure 10 in \citet{RomanDuvalEtAl2011} from the Galactic Ring Survey. Due to the stochastic nature of turbulent clouds we obtain a range of PCA slopes. We measure an average slope for the observable $^{13}$CO intensity of $0.74 \pm 0.02$ for the n300 model. This value is consistent with the values reported in \citet{RomanDuvalEtAl2011}: $0.62 \pm 0.2$ for a sample consisting of all clouds and including spatial scales greater than the FWHM of the telescope beam (48"). If only angular scales greater than twice the FWHM are included, \citet{RomanDuvalEtAl2011} obtain a mean slope of $0.74 \pm 0.3$. For the $^{12}$CO intensity we measure slopes $0.82 \pm 0.03$ for n300 and $0.59 \pm 0.02$ for n100, which are in reasonable agreement with the observational analysis by \citet{BruntAndHeyer2002b}, who find a mean value of $0.62 \pm 0.11$ for 23 fields in the FCRAO CO Survey of the Outer Galaxy.

Other studies also revealed a range of PCA slopes from 0.5 to 0.9. \citet{HeyerAndSchloerb1997} calculated principal components for a high spatial dynamic range of $^{12}$CO and $^{13}$CO data cubes of the Sh 155 (Cep OB3) cloud complex and found a range of indices from 0.42 to 0.55. \citet{HeyerAndBrunt2004} measured a slope of $0.65 \pm 0.01$ for $^{12}$CO $J=1 \rightarrow 0$ spectral imaging observations of 27 individual MCs. \citet{HeyerEtAl2006} analysed the Rosette and G216-2.5 clouds and inferred slopes of $0.74 \pm 0.04$ and $0.84 \pm 0.13$ for Rosette and $0.63 \pm 0.04$ and $0.56 \pm 0.02$ for G216-2.5 for $^{12}$CO and $^{13}$CO, respectively. \citet{FederrathEtAl2010} analysed high-resolution simulations of hydrodynamics and used PCA to compute slopes for purely solenoidal and compressive forcing and found values of $0.66 \pm 0.05$ and $0.76 \pm 0.09$, respectively. The slopes that we derived are consistent with the above studies and show that radiative transfer effects can cause variations in PCA slopes of $\approx \pm 0.1$, in agreement with previous studies \citep[see e.g.,][]{BruntAndHeyer2002a,BruntEtAl2003}.

\subsection{Turbulent intermittency effects}
\label{subsec:intermittency}

\citet{RomanDuvalEtAl2011} examined the dependence of the value of $\alpha_\text{PCA}$ on the density dispersion, finding that PCA is insensitive to the log-density dispersion $\sigma_s$, provided $\sigma_s \lesssim 2$. For $\sigma_s > 2$, $\alpha_\text{PCA}$ increases with $\sigma_s$ due to the intermittent sampling of the velocity field by the density field.

In order to characterise the density PDF and intermittency effects of the different simulations more quantitatively, it is convenient to define the quantity
\begin{equation}
s = \ln \left( \frac{n}{\langle n \rangle} \right),
\end{equation}
which is the natural logarithm of the density field $n$ divided by its mean value $\langle n \rangle$ \citep[see e.g.,][]{FederrathAndKlessen2012,FederrathAndKlessen2013}. We therefore compute the mean and standard deviation of $s$ for all models directly from the PPP cubes and list our results in Table \ref{tab:moments}. In order to focus only on the denser regions and to avoid a too strong influence from the very diffuse medium around, we restrict our computations on number densities above $0.05\,$cm$^{-3}$ for H$_2$ and the total density. For the total density in all models we find $\sigma_s < 2$ and hence expect the value of $\alpha_\text{PCA}$ to be reliable. For H$_2$ the density fields show a slightly broader dispersion that is consistent with limited chemical inhomogeneity, but still resulting in $\sigma_s < 2$. Hence, we also expect H$_2$ still to be described reliably \citep[see also][]{RomanDuvalEtAl2011}. On the other hand, we find a large dispersion in the CO density fields with $\sigma_s > 2$, which reflects a high degree of chemical inhomogeneity. Thus it is not surprising that we find different results for the $\alpha_\text{PCA}$ of CO compared to the total density models. It is also rather difficult to quantify intermittency of the observable CO intensity in a way such that it is comparable to the intermittency of the density fields. However, we expect optical depth effects to lead to smoothing of structures (see discussion in next section \ref{opticaldepth}) and hence to less intermittency. Therefore, we would expect this to yield $\alpha_\text{PCA}$ values closer to those from the underlying density field.

\begin{table}
\begin{tabular}{c|l|c|c|c}
\hline\hline
 Model name & Species & $\sigma_n / \langle n \rangle$ & $\sigma_s$ & $\langle s \rangle$ \\
\hline
 & Total density & 3.82 & 1.68 & -1.60 \\
 n30 & H$_2$ & 4.43 & 1.98 & -2.15 \\
 & CO & 57.67 & 5.18 & -15.88 \\
\hline
 & Total density & 2.54 & 1.45 & -1.11 \\
 n100 & H$_2$ & 3.22 & 1.74 & -1.49 \\
 & CO & 14.33 & 5.25 & -11.86 \\
\hline
 & Total density & 2.40 & 1.45 & -1.09 \\
 n300 & H$_2$ & 2.59 & 1.56 & -1.21 \\
 & CO & 4.20 & 5.75 & -6.78 \\
\hline\hline
\end{tabular}
\caption{Standard deviation $\sigma_n / \langle n \rangle$ of the density field $n$, log-density dispersion $\sigma_s$ and mean of $s$ for the latest evolved snapshot for each presented model n30, n100 and n300. For the total density and H$_2$ we find $\sigma_s < 2$ and hence expect the corresponding value of $\alpha_\text{PCA}$ to be reliable. The large dispersion of $\sigma_s > 2$ of the CO density fields can be refered to a high degree of chemical inhomogeneity.}
\label{tab:moments}
\end{table}

\subsection{Optical depth effects}
\label{opticaldepth}

\begin{figure}
\centerline{
\includegraphics[width=1.0\linewidth]{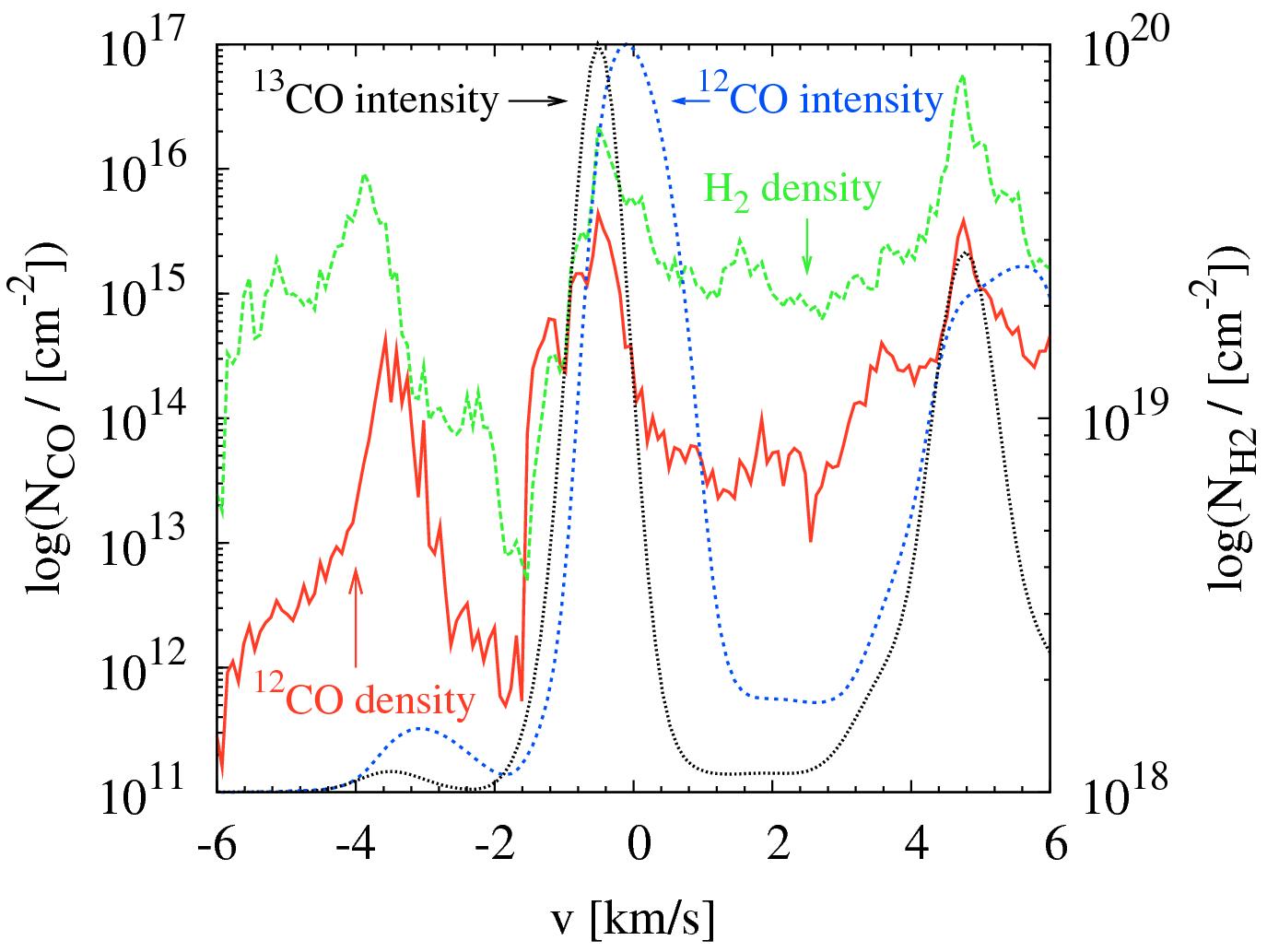} \\
}
\caption{Intensity and column density profiles along an arbitrary position in the model PPV cubes. Shown are the profiles for the $^{12}$CO and H$_2$ density and $^{12}$CO and $^{13}$CO intensity PPV cubes computed along the LoS in z-direction, averaged over a beam with radius of 10 pixels for the n100 model. The column density scales for the $^{12}$CO and H$_2$ density models are given on the left and right side, respectively. Both the $^{12}$CO and $^{13}$CO intensity are given in arbitrary units, normalized by the maximum value.}
\label{fig:spectra}
\end{figure}

It is important to understand the role of optical depth effects in the numerical simulations. Since $^{12}$CO is optically thick we may not be able to observe the higher density regions in the center of a core. This is because low-density regions toward the front of the LoS may ``hide'' the emission behind the low density gas, if both low and high density regions have the same velocity. In contrast, the $^{12}$CO density traces the ``true'' high density peaks since no optical depth effects are present. In the case of the $^{12}$CO density, PCA finds a lower degree of coherence on large scales compared to the $^{12}$CO intensity (Figure \ref{fig:PCA}). This is likely due to the fact that optical depth effects tend to smooth out the overall structure of the clouds.

Figure \ref{fig:spectra} shows the intensity and column density profiles along an arbitrary position in the model PPV cubes. Shown are the profiles for the $^{12}$CO and H$_2$ density and $^{12}$CO and $^{13}$CO intensity models computed along the LoS for the n100 model. We average over a beam with radius of 10 pixels (corresponding to 0.78~pc) in each velocity channel in PPV space. Comparing the spectra and maps of H$_2$ and $^{12}$CO density we find that the H$_2$ column density is more extended than the $^{12}$CO column density, with far more H$_{2}$ present in diffuse regions. Furthermore, we find that $^{12}$CO traces H$_2$ well only in the densest clumps, but not in low density regions. For the $^{12}$CO density case, PCA cannot detect structures on large scales since the contrast between the main density peaks and the diffuse regions around them is too high to observe coherent structures over larger scales.

As expected, the $^{12}$CO and $^{13}$CO intensity spectra are much smoother compared to the density spectra. In this case, PCA finds a high degree of coherence and therefore larger scale structures can be detected (see Figure \ref{fig:PCA}), since the contrast between the main peaks is not as high as in the density case. $^{12}$CO is optically thick which means that the $^{12}$CO density and intensity spectra do not scale linearly with each other. Furthermore, $^{13}$CO is optically thin and traces the H$_2$ gas very well.

Although the structure of the different fields (total density, H$_2$, $^{12}$CO density, $^{12}$CO and $^{13}$CO intensity) appear different in Figures \ref{fig:images} and \ref{fig:spectra}, the resulting PCA slopes are only marginally different (Table \ref{tab:PCAvalues}). Most notably, model n300 results in PCA slopes that are consistent within the 2$\sigma$ uncertainties between the $^{12}$CO, total and H$_2$, densities. In model n300, there are very high density peaks. In real clouds these would undoubtedly be sites of star formation, leading to higher UV radiation fields and the photodissociation of CO. These effects are not included in the simulation, and would certainly impact the PCA. For the fiducial n100 model, the optical depth effects and abundance variations lead to slightly discrepant PCA slopes when comparing different tracers, though the slopes are consistent within 4$\sigma$.

\section{Summary and Conclusions}
\label{sec:summary}

We applied PCA to numerical simulations of MCs with time dependent chemistry to describe the impact of chemistry, radiative transfer and of the mean density on the turbulent dynamics of MCs. We studied the PPV cubes of five components: the total number density, H$_2$, $^{12}$CO (each without radiative transfer) and $^{12}$CO and $^{13}$CO, respectively (each with radiative transfer). In each case, we analysed the PCA pseudo structure function for different initial number densities $n_0$. We computed individual slopes $\alpha_\text{PCA}$ and compared them to observational measurements. Furthermore, we also studied the influence of radiative transfer on the results of the PCA. We report the following findings:

\begin{enumerate}
\item $^{12}$CO is a good tracer of H$_2$ in compact regions in MCs, but not in low density regions. H$_2$ is more extended than the $^{12}$CO density as it is better able to self-shield in diffuse regions.
\item Optical depth effects can be important for the PCA structure analysis. Since $^{12}$CO is optically thick and can be easily excited, there is a much higher degree of coherence in the spatial dimension of MCs for the $^{12}$CO intensity than for the $^{12}$CO density, as shown in Section \ref{opticaldepth}. Large scales are therefore suppressed in the $^{12}$CO density. Since $^{13}$CO is generally optically thin, it traces the structure of a molecular clouds more accurately.
\item We are able to reproduce observations by performing simulations with chemistry and radiative transfer postprocessing for $^{12}$CO and $^{13}$CO, as shown in Figure \ref{fig:hist}. We measure slopes $\alpha_\text{PCA}$ of the pseudo structure function in a range from 0.5 to 0.9. The slope distributions in Figure \ref{fig:hist} peak at $0.59 \pm 0.02$ for $^{12}$CO intensity for the n100 and at $0.82 \pm 0.03$ and $0.74 \pm 0.02$ for the n300 model for $^{12}$CO and $^{13}$CO intensity, respectively.
\item We compute mean values of $\langle l_1 / l_0 \rangle \gtrsim 0.2$ that are consistent with previous studies of \citet{OssenkopfAndMacLow2002} and \citet{BruntEtAl2009} and show that the driving scale is not significantly affected by the chemistry and the radiative transfer.
\item We analyse the log-density dispersion $\sigma_s$ of the different PDFs, finding $\sigma_s < 2$ for the total density and H$_2$ density field. For the CO density field we find $\sigma_s > 2$ which is due to the high degree of chemical inhomogeneity.
\end{enumerate}

Despite the difference in the morphology evident in the maps (Figure \ref{fig:images} and \ref{fig:spectra}), most of the resulting PCA slopes for all components are only marginally different, depending on density. For the higher density model n300, where the effects from stellar feedback that are not modeled and are likely to affect the results, the turbulence statistics of $^{12}$CO and $^{13}$CO intensity provide reasonably good approximations to the underlying structure of the density and velocity field. For the fiducial n100 model, which has a lower density, differences between the underlying physical parameters and the emergent CO intensity lead to more notable discprencies in the corresponding $\alpha_\text{PCA}$.

For further investigations and to get a better understanding of how different initial conditions (e.g.\ different metallicities and chemical abundances, turbulent driving on different scales and amplitudes, different radiation fields, etc.) affect the slopes $\alpha_\text{PCA}$, we need numerical simulations that span a wider range of physical parameters. Varying the turbulence driving on different spatial scales could be necessary to model local feedback from other stars or from stellar winds.

\section*{Acknowledgements}

We thank Volker Ossenkopf, Nicholas Kevlahan, Lukas Konstandin and Mark Heyer for informative discussions about the project. We thank Chris Brunt for a timely, detailed and constructive referee report, which helped to improve the paper. EB, RS, SCOG and RSK acknowledge support from the Deutsche Forschungsgemeinschaft (DFG) via the SFB 881 (sub-projects B1, B2 and B5) ``The Milky Way System'', and the SPP (priority program) 1573, ``Physics of the ISM''. C.F.~acknowledges funding provided by the Australian Research Council under the Discovery Projects scheme (grant no.~DP110102191). The simulations analysed in this paper were performed using the Ranger cluster at the Texas Advanced Computing Center, using time allocated as part of Teragrid project TG-MCA99S024.

\begin{appendix}

\section{Resolution study}
\label{app:resol}

\begin{table}
\begin{tabular}{|c|c|c|c}
\hline\hline
 Res & Total density & H$_2$ & $^{12}$CO \\
\hline
 $128^3$ & $0.86 \pm 0.06$ & $0.89 \pm 0.07$ & $0.89 \pm 0.09$\\
 $256^3$ & $0.77 \pm 0.03$ & $0.78 \pm 0.03$ & $0.89 \pm 0.05$\\
 $512^3$ & $0.63 \pm 0.03$ & $0.63 \pm 0.03$ & $0.64 \pm 0.04$\\
\hline\hline
\end{tabular}
\caption{Slopes $\alpha_\text{PCA}$ for different numerical models without radiative transfer for a resolution of 128$^3$, 256$^3$ and 512$^3$.}
\label{tab:resolution}
\end{table}

\begin{figure}
\centerline{
\includegraphics[width=1.0\linewidth,height=0.61\linewidth]{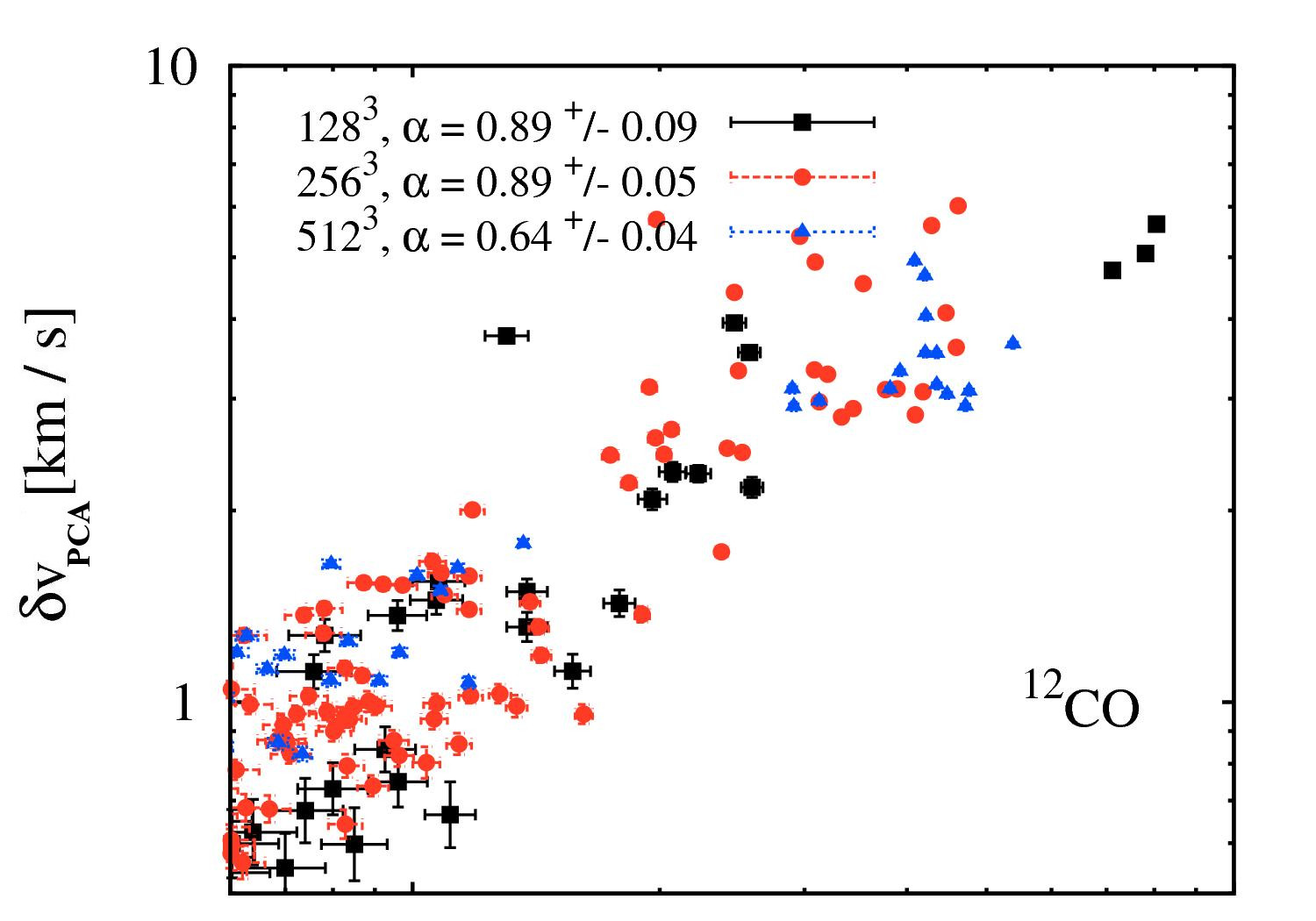} \\
}
\centerline{
\includegraphics[width=1.0\linewidth]{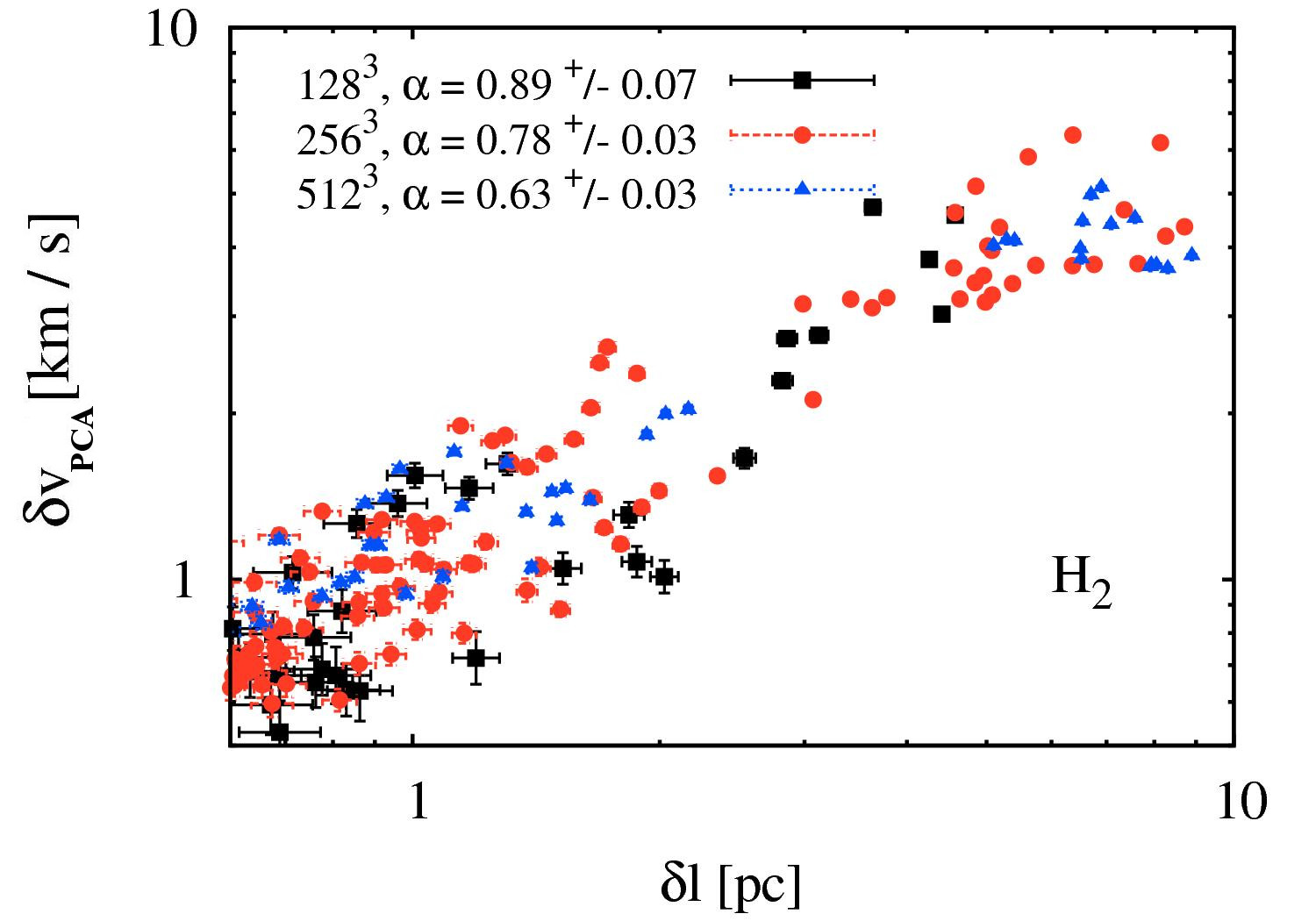} \\
}
\caption{Same as Figure \ref{fig:PCA}, but for different resolutions of 128$^3$, 256$^3$ and 512$^3$ grid cells for the n300 model for $^{12}$CO and H$_2$. The scatter of points decreases with increasing resolution.}
\label{fig:resolution}
\end{figure}

Figure \ref{fig:resolution} shows the results for the n300 model (see Appendix \ref{app:realizations}) for $^{12}$CO and H$_2$ for a resolution of 128$^3$, 256$^3$ and 512$^3$ grid cells. For all resolution models we observe a large scatter of points in $\delta v_\text{PCA}$, which decreases with increasing resolution. Table \ref{tab:resolution} gives the slopes for the resolution study. We clearly measure an effect of resolution dependence since the slopes decrease with increasing resolution for all chemical models. Previous studies already revealed that at scales smaller than 30 pixels for simulations of 256$^3$ grid cells numerical resolution effects influence the physical quantities \citep{KitsionasEtAl2009,FederrathEtAl2010,FederrathEtAl2011,SurEtAl2010}, which means that our exponents are not numerically converged. However, we do not consider the true PCA slope, but rather analyse the influence of a radiative transfer post-processing on the chemical models. We therefore compare PCA slopes of individual species with one another. Nevertheless, we note that future efforts that focus on measuring the intrinsic PCA slopes have to consider high resolution simulations with a chemical modeling and a radiative transfer post-processing.

\section{Use of different realizations of the simulations}
\label{app:realizations}

\begin{figure}
\centerline{
\includegraphics[width=1.0\linewidth]{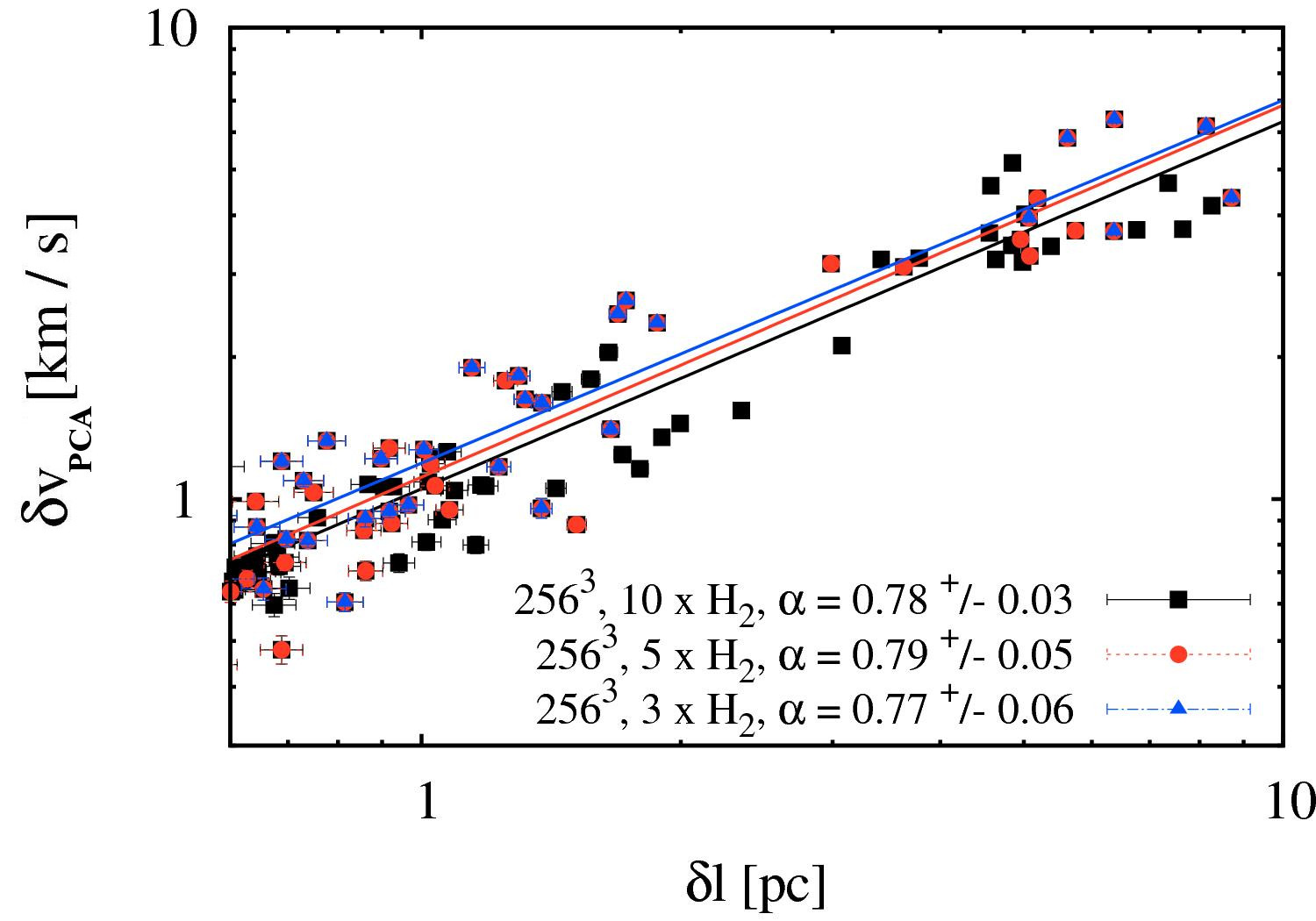} \\
}
\caption{Same as Figure \ref{fig:resolution}, but with fits for 10, 5 and 3 different realizations for the H$_2$ model of n300. All slopes are consistent with each other within their errors.}
\label{fig:seeds}
\end{figure}

For the n300 model we use ten different realizations with different random seeds for the initial turbulent velocity field. Each run is evolved until the same time where both the dynamical and chemical properties of the simulations have reached a steady state. Figure \ref{fig:seeds} shows fits for 10, 5 and 3 different realizations for the n300 model. The PCA slopes $\alpha_\text{PCA}$ with values of $0.78 \pm 0.03$, $0.79 \pm 0.05$ and $0.77 \pm 0.06$ are consistent with each other within their errors which means that already for 3 realizations (with 3 LoS directions) we operate in a converged slope regime. It is thus not necessary to use a larger number of snapshots to obtain a stable and robust slope.

\bibliographystyle{mn2e}

\end{appendix}
\end{document}